\documentclass{aastex}
\usepackage{natbib,epsfig,emulateapj5,rotate}

\tightenlines

\begin{document}
\newcommand{\be}{\begin{eqnarray}}
\newcommand{\ee}{\end{eqnarray}}
\newcommand{\hii}{\ion{H}{2}}
\newcommand{\hi}{\ion{H}{1}}
\newcommand{\etal}{et al.}
\def\llbdm{{\cal L}_{\rm \ell,b,\DM}}
\def\lpsrd{{\cal L}_{\rm psr,d}}
\def\lpsrs{{\cal L}_{\rm psr,s}}
\def\lgals{{\cal L}_{\rm gal,s}}
\def\lxgals{{\cal L}_{\rm xgal,s}}
\def\like{{\cal L}}
\newcommand{\lampsrd}{{\Lambda}_{\rm psr,d}}
\newcommand{\lampsrs}{{\Lambda}_{\rm psr,s}}
\newcommand{\lamgals}{{\Lambda}_{\rm gal,s}}
\newcommand{\lamxgals}{{\Lambda}_{\rm xgal,s}}
\def\lam{{\Lambda}}
\def\dl{{\rm D_L}}
\def\du{{\rm D_U}}
\newcommand{\Deff}{D_{\rm eff}}
\def\norm{{\cal N}}
\def\npsrd{{N_{\rm psr,d}}}
\def\dhat{{\hat D}}
\def\DM{{\rm DM}}
\def\RM{{\rm RM}}
\def\SM{{\rm SM}}
\def\EM{{\rm EM}}
\def\DMinfty{{\rm DM}_{\infty}}
\newcommand{\SMc}{\SM_c}
\newcommand{\DMc}{\DM_c}

\def\SMxgal{{\rm SM}_{\theta,x}}
\def\SMgal{{\rm SM}_{\theta,g}}
\def\SMtau{{\rm SM}_{\tau}}

\def\smun{{kpc \,\, m^{-20/3}}} 
\def\cnsq{{C_n^2}} 
\def\nbar{{\overline{n}_e}}
\def\dne{{\delta n_e}}
\def\pne{{P_{\delta \rm n_e}}}
\newcommand{\dphi}{D_{\phi}}

\def\dnud{{\Delta\nu_{\rm d}}}
\def\taud{{\tau_d}}
\def\narms{{N_{\rm arms}}}
\def\nclumps{{N_{\rm clumps}}}
\def\nvoids{{N_{\rm voids}}}

\def\nparms{{N_{\rm parms}}}
\def\nmc{{N_{\rm MC}}}

\def\xvec{{\bf x}}
\newcommand{\xbar}{\overline {x}}
\newcommand{\ybar}{\overline {y}}
\newcommand{\zbar}{\overline {z}}

\newcommand{\rc}{r_{\rm c}}
\newcommand{\dc}{d_{\rm c}}
\newcommand{\nec}{n_{\rm e_c}}
\newcommand{\Fc}{F_{\rm c}}

\newcommand{\sgra}{\mbox{Sgr~A${}^*$}}
\newcommand{\dgc}{\ensuremath{D_{\mathrm{GC}}}}
\newcommand{\delgc}{\ensuremath{\Delta_{\mathrm{GC}}}}
\newcommand{\rsun}{\ensuremath{\mathrm{R}_\odot}}
\newcommand{\wlism}{\ensuremath{w_{\rm lism}}}
\newcommand{\wlhb}{\ensuremath{w_{\rm lhb}}}
\newcommand{\wlsb}{\ensuremath{w_{\rm lsb}}}
\newcommand{\wldr}{\ensuremath{w_{\rm ldr}}}
\newcommand{\wloopI}{\ensuremath{w_{\rm loop I}}}
\newcommand{\wvoids}{\ensuremath{w_{\rm voids}}}
\newcommand{\halpha}{\ensuremath{\rm H\alpha}}
\newcommand{\nelism}{n_{\rm lism}}
\newcommand{\neldr}{n_{\rm ldr}}
\newcommand{\nelsb}{n_{\rm lsb}}
\newcommand{\nelhb}{n_{\rm lhb}}
\newcommand{\neloopI}{n_{\rm loop I}}
\newcommand{\nevoids}{n_{\rm voids}}
\newcommand{\neclumps}{n_{\rm clumps}}
\newcommand{\Flhb}{F_{\rm lhb}}
\newcommand{\Fldr}{F_{\rm ldr}}
\newcommand{\Flsb}{F_{\rm lsb}}
\newcommand{\FloopI}{F_{\rm loop I}}
\newcommand{\Flism}{F_{\rm lism}}
\newcommand{\negal}{n_{\rm gal}}
\newcommand{\negc}{n_{\rm GC}}
\newcommand{\nearms}{n_{\rm arms}}
\newcommand{\ok}{$\surd$}
\newcommand{\bvec}{{\bf b}}
\newcommand{\nutrans}{\nu_{\rm trans}}
\newcommand{\nover}{N_{\rm over}}
\newcommand{\nhits}{N_{\rm hits}}
\newcommand{\nlum}{N_{\rm lum}}
\newcommand{\smin}{S_{\rm min}}
\newcommand{\Dmax}{D_{\rm max}}
\newcommand{\Lp}{L_{\rm p}}
\newcommand{\lp}{l_{\rm p}}
\newcommand{\fLp}{f_{\Lp}}
\newcommand{\flp}{f_{\lp}}
\newcommand{\fLpP}{f_{\lp, P}}
\newcommand{\fLpDM}{f_{\Lp, \DM}}
\newcommand{\flpDM}{f_{\lp, \DM}}
\newcommand{\fDM}{f_{\DM}}
\newcommand{\fD}{f_{\rm D}}
\newcommand{\RGC}{R_{\mathrm{GC}}}
\newcommand{\HGC}{H_{\mathrm{GC}}}
\def\data{{\cal D}}
\def\models{{\cal M}}
\def\info{{\cal I}}
\newcommand{\thetavec}{{ \mbox{\boldmath $\theta$} }}
\newcommand{\rperp}{r_{\perp}}
\newcommand{\cu}{C_u}
\newcommand{\csm}{C_{\rm SM}}
\newcommand{\rscatt}{r_{\rm scatt}}


\title{ 
NE2001. I.  A New Model for the Galactic Distribution\\
  of Free Electrons and its Fluctuations}
\author{J. M. Cordes}
\affil{Astronomy Department and NAIC, Cornell University,
 Ithaca, NY~~14853\\ cordes@spacenet.tn.cornell.edu}
\bigskip
\author{T.~Joseph~W.~Lazio}
\affil{Naval Research Lab, Code~7213, Washington, D.C. 20375-5351  \\
Joseph.Lazio@nrl.navy.mil}

\bigskip
\begin{abstract}
We present a new model for the Galactic distribution of free
electrons.  
It 
(a) describes the distribution of the free electrons responsible 
for pulsar dispersion measures and thus can be used for estimating 
the distances to pulsars;
(b) describes large-scale variations in the strength
of fluctuations in electron density that underly interstellar scattering;
(c) can be used to interpret interstellar scattering and scintillation
observations of Galactic objects and of extragalactic objects,
such as intrinsically compact AGNs  and Gamma-ray burst afterglows;
and
(d) serves as a preliminary, smooth spatial model of the 
warm ionized component of the interstellar gas.
This work builds upon and supersedes the Taylor \& Cordes (1993)
model by exploiting new observations and methods, 
including 
(1)~a near doubling in the number of lines of sight
   with dispersion measure or scattering measurements; 
(2)~a substantial increase in the number and quality of independent
   distance measurements or constraints;
(3)~improved constraints on the strength and
   distribution of scattering in the Galactic center; 
(4)~improved constraints on the (Galactocentric) radial distribution of free
    electrons; 
(5) redefinition of the Galaxy's spiral arms, including the influence
    of a local arm;
(6) modeling of the local interstellar medium, including the local hot
bubble identified in X-ray and \ion{Na}{1} absorption measurements;
and 
(7)~an improved likelihood analysis for
   constraining the model parameters.
For lines of sight directed out of the Galactic plane, the new model
yields substantially larger values for pulsar dispersion measures,
expect for directions dominated by the local hot bubble.
Unlike the TC93 model, the new model provides sufficient electrons to
account for the dispersion measures  of the vast majority of
known,  Galactic pulsars.
The new model is described and exemplified using plots of astronomically
useful quantities on Galactic-coordinate grids.  Software available on
the Internet is also described. 
Future observations and analysis techniques
 that will improve the Galactic model are outlined. 
\end{abstract}

\section{Introduction}\label{sec:intro}

Radio wave propagation measurements provide unique information about
the magnetoionic component of the interstellar medium (ISM).  Pulsars
are important probes because they emit short duration radio pulses
that are modified by intervening plasma and also because they are
highly spatially coherent, allowing scattering processes to significantly
perturb their radiation.    Other compact sources, 
both Galactic and extragalactic, also serve as probes of the plasma.      

In this first of a series of papers, we present a new model for the
free electron density of the Galaxy.  It is called ``NE2001'' because
it incorporates data obtained or published through the end of~2001.  A
short history of such models prior to~1993 is given in Taylor \&
Cordes~(1993, hereafter TC93).  Our work builds upon and supersedes
the TC93 model by exploiting new dispersion and scattering
measurements and also by employing new techniques for modeling.
Following Cordes \etal~(1991) and TC93, we model fluctuations in
electron density as well as the local mean density.
 
Since TC93 was written, significant developments have taken place
that increase the sample of relevant measurements and indicate
shortcomings of the model.  
Most importantly,
independent distance measurements are available on about 50\% 	more 
objects.   Some of these are precise parallax measurements obtained
through pulse timing or interferometric techniques.  Others result
from \ion{H}{1} absorption measurements combined with a kinematic rotation
model for the Galaxy.  Still others arise from the association
of pulsars with supernova remnants and globular clusters.  Distance
estimates combined with dispersion measures quantify the line-of-sight
average electron density.    
The new distance constraints indicate, in a minority of cases,
  that some of the distances derived from the TC93 model using pulsar 
dispersion measures are in error by factor of two or more.  However,
we point out that some of the parallax measurements 
also differ from previous ones by significant amounts.  
Additional information
is provided by the distribution of dispersion measures for the entirety
of available pulsar samples. The number of
such measurements is about double of that available in 1993.
Constraints on the square of the electron 
density arise from scattering measurements such as angular
broadening, pulse broadening and diffractive scintillation measurements.
The number of
scattering measurements has almost doubled since 1993. 

The TC93 model is flawed in several respects, some of which were
apparent even at the time of its development.  First, it provides
insufficient column density at high Galactic latitudes so that only
lower bounds on pulsar distances can be derived from it.  Second, in
some directions, particularly in the fourth quadrant along tangents to
the Carina-Sagittarius and Crux-Scutum spiral arms, the model provides
either too many electrons for some objects (Johnston \etal~2001) or
too few for others (as discussed below), indicating that the spiral
arms need redefinition.  Third, in the direction of the Gum Nebula and
Vela supernova remnant, the TC93 model provides too little scattering
to account for the pulse broadening of some pulsars (Mitra \&
Ramachandran 2001).  Fourth, interstellar scintillations of nearby
pulsars have scintillation bandwidths (which measure the column
density of electron density fluctuations) that are not well modeled
(Bhat \& Gupta~2002).  Fifth, similar to the first deficiency, the
calibration between scattering of Galactic and extragalactic sources
needs revision in order to ascertain the column density of scattering
material toward cosmological sources as compared to that of Galactic
objects outside of but near the apparent boundary of free electrons.

Our knowledge of the ISM has increased significantly from a host of
other investigations across the electromagnetic spectrum.
The local ISM, in particular, is now much better modeled and we
incorporate that information into our model for the free electron density.
The local ISM has been probed using continuum  X-ray measurements and 
absorption of \ion{Na}{1} toward nearby stars.

Other models for the mean electron density have been presented since
TC93.  These include the recent work of 
G\'omez, Benjamin \& Cox (2001; hereafter GBC01), who presented
a two-component, axisymmetric model, not dissimilar in form to
that used by Cordes \etal\ (1991). 
The two components have 
sech$^2(r/R)$sech$^2(z/H)$ variations with different
radial and $z$ scales.  
GBC01 do not attempt to model electron density
fluctuations relevant to scattering and their model is based solely
on the 109 objects available to them that have independent distance
estimates.  As we demonstrate, though their model is adequate for nearby pulsars,
it grossly underpredicts the dispersion measures of many distant pulsars
at low Galactic latitudes  and some at high latitudes.  Their work
underscores the need, demonstrated before by others 
(Ghosh \& Rao 1992; TC93), for spiral-arm structure in the free-electron
distribution.  

Our model uses significantly larger data samples than were available for
TC93 and makes use of all available data, including independent
distance constraints on pulsar distances, dispersion and
scattering data on pulsars, and scattering of other Galactic as well
as extragalactic sources.   We also  incorporate published, multiwavelength
data that allows modeling of the local ISM and of the spiral arms of
the Galaxy.


In this first paper we present the model and its usage.
In a second paper (Cordes \& Lazio 2002b; hereinafter Paper~II) 
we describe the input data and associated references, 
the likelihood analysis and modeling, alternative model
possibilities, and discussions of particular lines of sight.  Future
papers will apply the model to various astronomical and astrophysical
applications.  The plan of this paper is as follows.  
In \S\ref{sec:observables} we
   describe the observable quantities that we use to constrain the NE2001
   model parameters. 
In \S\ref{sec:model} 
   we describe the various components of the
   NE2001 model.
In \S\ref{sec:performance} we demonstrate the model's ability to
account for the distances and scattering of pulsars and discuss
briefly astronomical applications of the model.  Extensive discussion 
of applications of NE2001 is deferred to a later paper.

In \S\ref{sec:discussion} we summarize the results and outline future
prospects for improving the model.  These will rest largely on
usage of the Parkes Multibeam Pulsar sample (e.g. Manchester \etal\ 2001), 
improved parallax measurements using very long baseline interferometry,
and incorporation of additional multiwavelength observations into
the model definition and fitting.
Appendix~\ref{app:LISM} describes our model for the local interstellar
  medium.
Appendix~\ref{app:fortran} describes how to obtain the model as a set
of Fortran subroutines and its implementation in tools available through
the World Wide Web.

\section{Observables} \label{sec:observables}

In constructing the new model, we have used a variety of 
measurements, largely at radio wavelengths, but also including
the results of optical and X-ray observations that probe various
aspects of the local \hbox{ISM}.
In this  section we summarize the measurements used, defining 
line-of-sight integrated quantities in some detail because
we provide expressions and software  for estimating them using our
model. 

Radio data
consist of measurements along particular lines of sight
of propagation effects
that are sensitive to the electron density and its fluctuations.
In addition, there are independent distance estimates or constraints
based on a variety of techniques (\ion{H}{1} absorption, interferometric or
timing parallaxes, and associations with globular clusters or
supernova remnants).
The wave-propagation data include: 
(a) dispersion measures (DMs) of pulsars  obtained through measurements of
    differential arrival times;
(b) temporal broadening of pulses from pulsars with large \DM\
    caused by multipath scattering
    from electron density variations, $\dne$;
(c) scintillation bandwidth measurements of low-\DM\
    pulsars;
(d) angular broadening of Galactic and extragalactic sources caused
    by scattering from $\dne$;
and
(e) emission measures.

In Table \ref{tab:summary} we summarize the data sets by 
specifying the numbers of each kind of measurement.  
Figure~\ref{fig:venn} shows the relative number of measurements
for pulsars. 

\smallskip
\epsfxsize=8truecm
\epsfbox{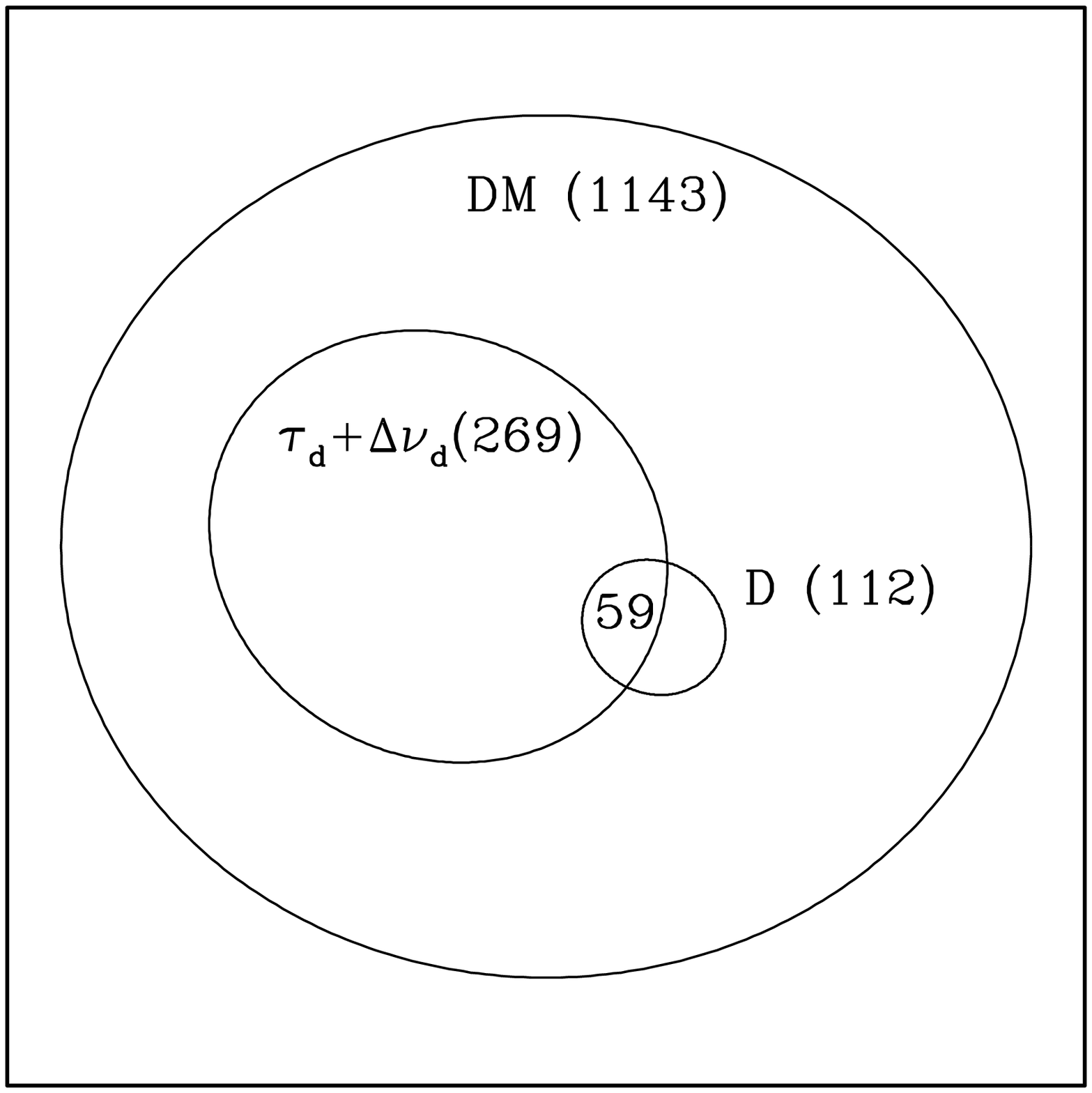}
\figcaption{
\label{fig:venn}
Venn diagram showing the number of DM measurements, independent
distance measurements (D), scattering measurements
(pulse broadening ($\taud$) + scintillation bandwidth ($\dnud$) measurements), 
and the overlap of pulsar
distance and scattering measurements.  In addition to 
pulsar data, there are 118 angular broadening measurements
on non-pulsar Galactic sources and on 94 extragalactic sources.
These numbers include upper limits.  See Table~\ref{tab:summary}
for additional information.
}

\subsection{\DM, \EM, and Distance Measurements}

The data used include 1143 dispersion measures, 
\begin{equation}
{\rm DM}\equiv\int_0^D ds\, n_e, \label{eq:dm}
\end{equation}
(Taylor, Manchester, \& Lyne 1993; 
Princeton pulsar catalog;
\footnote{http://pulsar.princeton.edu/pulsar/catalog.shtml}
Parkes Multibeam survey
\footnote{http://www.atnf.csiro.au/research/pulsar/pmsurv/}
).

We characterize distance constraints (e.g., from \ion{H}{1} absorption 
measurements) with an interval
$[\dl, \du]$, where the lower and upper distance limits, $\dl,\du$, 
are assumed to be hard limits, with uniform probability within the interval.
A compilation with references is given 
on a web site.
\footnote{http://www.astro.cornell.edu/$\sim$shami/psrvlb/parallax.html}
The best distance constraints are from parallax measurements
using pulse timing techniques (e.g. van Straten \etal\ 2001)
and very-long-baseline interferometry 
(Chatterjee \etal\ 2001; Brisken \etal 2002).

The emission measure \EM\ is the integral of the squared electron density,  
\be
\EM = \int_0^D ds \, n_e^2, \label{eq:em}
\ee
and thus provides information about fluctuations in electron density
and their spatial distribution.  

\subsection{Radio Wave Scattering Measurements}

We use three kinds of measurements: the angular broadening
(``seeing'') diameter $\theta_d$; the pulse broadening time, $\taud$,
that measures the temporal smearing associated with the multipath
propagation; and the scintillation bandwidth, $\dnud$, the frequency
range over which diffractive scintillations are correlated.  All three
observables are sensitive to the detailed distribution along the line
of sight of the scattering strength, owing to standard geometrical
leveraging effects encountered in optics.  Details are summarized
below and discussed thoroughly in Cordes \& Rickett (1998, hereafter
CR98; see also Deshpande \& Ramachadran 1998).

Angular broadening is
characterized as the full-width at half-maximum (FWHM) of the 
scattering contribution to the measured brightness distribution.

The pulse broadening time $\taud$ is determined by deconvolving
intrinsic and measured pulse shapes.  The broadening
times are roughly the $e^{-1}$ time
of the pulse broadening function, which is roughly a
one-sided exponential 
(e.g., Cordes \& Rickett 1998; Lambert \& Rickett 1999).

The scintillation bandwidth ($\dnud$) is estimated 
as the half-width at half maximum of the 
autocorrelation function of intensity  along the 
frequency-lag axis
(Cordes 1986; Johnston, Nicastro, \& Koribalski 1998; 
Bhat, Rao, \& Gupta 1999).
The ``uncertainty'' relation 
between $\taud$ and $\dnud$ may be written
as $2\pi\dnud\taud = C_1$, where $C_1$ is a constant
that depends on the wavenumber spectrum and on the large-scale
distribution of scattering material along the line of sight
(CR98).  
We use $C_1 = 1.16$, 
the value appropriate for a uniform medium
and having a Kolmogorov wavenumber spectrum.

\subsubsection{Density Spectrum and Scattering Measure}

We assume that 
$\dne\propto n_e$ but with a proportionality constant
that varies between different components of the Galactic plasma.
Also, we adopt a power-law wavenumber spectrum for $\dne$,
\be
\pne(q) = \cnsq\, q^{-\beta}, \quad 
	\frac{2\pi}{\ell_0} \le q \le \frac{2\pi}{\ell_1};
\ee
$\ell_1$ and $\ell_0$ are the inner and outer scales 
	of the fluctuations in $\dne$ and
$\cnsq$ is the spectral coefficient (the ``level of turbulence'').  
We take $\beta = 11/3$, the ``Kolmogorov'' spectrum because it is
appropriate for many lines of sight 
(e.g. Lee \& Jokipii 1976; Armstrong, Rickett \& Spangler 1995).
We point out that there are numerous caveats on this choice of
$\beta$, both empirical and theoretical, which we discuss in
Paper II.\footnote{Briefly, some observable quantities scale with
frequency or distance differently than what is expected for a
simple Kolmogorov model.    Estimates of \SM\ based on measured
quantities may accordingly be in error.   However, these errors
are no more than a factor of a few and can be compared with
line-of-sight variations of several orders of magnitude in \SM\
that we aim to model.}
The scattering measure is the path integral of $\cnsq$:
\be
{\rm SM} \equiv \int_0^D ds\, C_n^2. \label{eq:smdef}
\ee

Different observables correspond to different LOS weightings for
$\cnsq$, yielding  different effective values for the scattering measure.
Following Cordes \etal\ 1991 and TC93,  we define expressions for
the {\it effective} \SM\ if a uniform medium is assumed:
\begin{list}{}
\item 
(1) For measurements of angular diameters of extragalactic sources,
\be
	\SMxgal = \SM. \label{eq:smxgal} 
\ee
\item 
(2) For angular diameters of Galactic sources,
\be
\SMgal = 3\int_0^D ds\, (1-s/D)^2\cnsq, 
\label{eq:smgal}
\ee
 where the integration is from observer to source.
\item 
(3) For pulse broadening and scintillation measurements,
\be
\SMtau = 6\int_0^D ds\, (s/D)(1-s/D)\cnsq. 
\label{eq:smpsr}
\ee
\end{list}
With these expressions,   the true \SM\ for a statistically  uniform medium
would be estimated correctly.   For nonuniform media, the results
are only approximations to the true \SM.

\medskip
Under the same assumption of a uniform medium, 
we calculate observables in terms of these weighted SMs 
using (with $\nu$ in GHz and \SM\, in units defined above):
\be
\theta_d = 
 \nu^{-11/5} \times
\left\{
\begin{array}{ll}
128 \,{\rm mas}\,\, \SMxgal^{3/5} 
	& \quad\mbox{\rm Extragalactic Sources} \\
\\
~71 \,{\rm mas}\,\, \SMgal^{3/5}  
	& \quad\mbox{\rm Galactic Sources}. \\
\end{array}
\label{eq:thetad}
\right.
\ee
The pulse broadening time  and scintillation bandwidth are given by 
\be
\taud &=& 1.10\, {\rm ms}\,\,   \SMtau^{6/5}\nu^{-22/5}D \label{eq:taud}\\
\nonumber \\
\dnud &=& 168\, {\rm Hz} \,\,   \SMtau^{-6/5}\nu^{22/5}(C_1/1.16) D^{-1}. 
\label{eq:dnud}
\ee
The coefficient in Eq.~\ref{eq:dnud} is the same as Eq.~48 of 
Cordes \& Lazio (1991) if, as assumed there, $C_1 = 1.53$, the value
appropriate for a medium with a square-law structure function.
However, $C_1 = 1.16$ applies for a Kolmogorov medium that is 
statistically homogeneous and which we consider to be
a better default model than  a medium with square-law structure function.

Following Cordes \etal\ (1991) and TC93, we relate $\cnsq$ to
the local mean electron density inside ionized clouds 
$\nbar$ and calculate
line of sight integrals taking into account the volume filling factor
$\eta$ of those clouds and cloud-to-cloud variations of the internal density.
We use $n_e = \eta\nbar$ to denote the local spatially-averaged density.
We relate \SM\, to \DM\, using
\be
d\,{\rm DM} &=& n_e\,ds, \label{eq:DM} \\
d\,{\rm SM} &=& \csm~F~n_e^2\, ds\,, \label{eq:SM} 
\ee
where $F$ is a fluctuation parameter,
\be
F = \zeta\epsilon^2\eta^{-1} \ell_0^{-2/3}\ ,
\label{eq:F}
\ee
that depends on 
the fractional variance inside clouds, 
$\epsilon^2= \langle (\delta {n_e})^2\rangle/\nbar^2$, 
the normalized second moment of cloud-to-cloud fluctuations in $\nbar$,
$\zeta = \langle \nbar^2 \rangle / \langle \nbar \rangle^2$, 
and the outer scale $\ell_0$ expressed in parsec units.
We define the constant, $\csm = [3(2\pi)^{1/3}]^{-1}\cu$, 
where the scale factor $\cu = 10.2$~m$^{-20/3}$~cm$^6$
yields SM in the (unfortunately) conventional units of kpc~m$^{-20/3}$ for 
$n_e$ in cm$^{-3}$, $ds$ in kpc, and $\ell_0$ in \hbox{AU}.

The emission measure may be expressed as
\be
d\EM\ &=& [3(2\pi)^{1/3}] \ell_0^{2/3} 
	\epsilon^{-2}(1+\epsilon^2)\,d\SM\ \nonumber \\
     &=& 544.6\,\, {\rm pc\,\,cm^{-6}}\,  \ell_0^{2/3} 
		\epsilon^{-2} (1+\epsilon^2)\,d\SM.
\label{eq:sm2em}
\ee

For completeness, we provide expressions for the free-free optical depth,
the related intensity of H$\alpha$ emission, and the transition frequency
between weak and strong scattering. These quantities are evaluated
using subroutines provided in the software described in
Appendix~\ref{app:fortran}.
The free-free optical depth is 
$\tau_{\rm ff} = 5.47 \times 10^{-8}\,\nu^{-2}\,T_4^{-3/2}\,g(\nu, T)\,\mathrm{EM}$
($\nu$ in GHz, $T_4 =$ temperature in units of $10^4$ K, and $g =$ Gaunt
factor $\sim 1$) (Rybicki \& Lightman 1979, p. 162).
Relating EM to SM, we find that 
the implied free-free optical depth is
\be 
\tau_{\rm ff}  = 10^{-4.53}\ \nu^{-2}\ T_4^{-3/2} \nu^{-2} 
	 \ell_0^{2/3}\  \epsilon^{-2} (1+\epsilon^2) \SM.
\label{eq:tau_ff}
\ee
The intensity of $\halpha$ emission (in Rayleighs) in terms of \SM\ 
(Haffner, Reynolds, \& Tufte 1998, Eq.~1) is
\be
I_{\halpha} = 198 R\, T_4^{-0.9} \epsilon^{-2} (1 + \epsilon^2) \ell_0^{2/3}
	\SM.
\ee

The expression for the transition frequency between weak and strong
scattering (Rickett 1990) used in the software discussed in
Appendix~\ref{app:fortran} is
\be
\nutrans = 318\, {\rm GHz}\, \xi^{10/17} \SM^{6/17} \Deff^{5/17},  
\label{eq:nutrans}
\ee
where $\Deff$ is now an effective distance to the scattering medium 
and the factor $\xi \sim 1$ allows scaling of one's preferred definition
for the Fresnel scale (e.g. $\xi = (2\pi)^{-1/2}$ is commonly used,
decreasing the coefficient by a factor 0.58).
For spherical waves embedded in a medium with constant $\cnsq$, 
the coefficient is multiplied by a factor
$(\beta -1)^{-2/(\beta+2)}$ and becomes 225 GHz.   
Thus, for nearby pulsars with 
$\log \SM \approx -4$ and $D\approx 0.1$ kpc, $\nutrans \approx 5$ GHz.

\section{Galactic Model for Electron Density}\label{sec:model}

\subsection{Basic Structure}\label{sec:basics}

As in TC93, we use a right-handed coordinate system 
$\xvec = (x,y,z)$
with its origin
at the Galactic center, $x$ axis directed parallel to $l=90^\circ$,
and $y$ axis pointed toward $l=180^\circ$.  
The Galactocentric distance projected onto the plane  is
$r=(x^2+y^2)^{1/2}$. 

The electron density is the sum of two axisymmetric components
and a spiral arm component, combined with terms that describe
specific regions in the Galaxy.   Table~\ref{tab:ne2001} gives the
details of the functional forms, which we describe briefly here. 

%

We have distinguished terms that represent the local ISM ($\nelism$),
the large scale distribution ($\negal$), the Galactic center
($\negc$), and individual clumps (${\neclumps}$) and voids (${\nevoids}$).
The large scale distribution, $\negal$, consists
of two axisymmetric components,  a thick (1) and  thin (2) disk,
and spiral arms. 
The weight factor 
$\wlism = 0,1$ switches off or on the local ISM
component and switches on or off the smooth, large scale
components of the model (and also the Galactic-center component, $n_{\rm GC}$).
Superposed with the large-scale and local-ISM components are
``clumps'' of excess
electron density that we infer from the database of measurements as
outliers from the smooth model.  They most likely correspond to individual
\ion{H}{2} regions or portions of ionized shells surrounding supershell regions.  
Finally, we also include ``voids'' that generally are regions
of lower than ambient density which are mutually exclusive of all 
components (except clumps) rather than superposing with them.  
Voids override all components other than clumps
and are required to account for the distances of some pulsars.

Associated with each component in the model 
is a separate value of the fluctuation parameter, $F$. 
Details about the functions used 
are given in  Table~\ref{tab:ne2001}  and summarized in sections
below.
As in TC93, we use  ${\rm sech}^2(\vert z\vert /H )$ 
for the $z$ dependences of most components to
produce a ``rounder'' variation at $z=0$ than an exponential
dependence $\propto \exp(-\vert z\vert/H)$.  
Both the exponential and sech$^2$ functions integrate to
the same asymptotic value ($H$) and have nearly equal $1/e$ locations.

Table~\ref{tab:large} gives parameter values for the large scale
parameters and compares them, where appropriate, with those of the
TC93 model.  The best-fitting parameters were found by an iterative
likelihood analysis that is similar to that used in TC93 but with a
number of improvements in the details of the fitting (Paper~II).

\medskip
\epsfxsize=8truecm
\epsfbox[107 192 477 562]{figI.2.ps}
\figcaption{
\label{fig:grayscale}
Electron density corresponding to the best
fit model  plotted as a grayscale with logarithmic levels on
a 30$\times 30$ kpc x-y plane at z=0 and centered on the
Galactic center.  The most prominent large-scale features
are the spiral arms, a thick, tenuous disk, a molecular ring
component. A Galactic center component  appears as a small dot.
The small-scale, lighter features represent
the local ISM and underdense regions required for by some lines of sight with
independent distance measurements.  The small dark region  embedded
in one of the underdense, ellipsoidal regions is the Gum Nebula and
Vela supernova remnant. 
}

\medskip
\epsfxsize=8truecm
\epsfbox{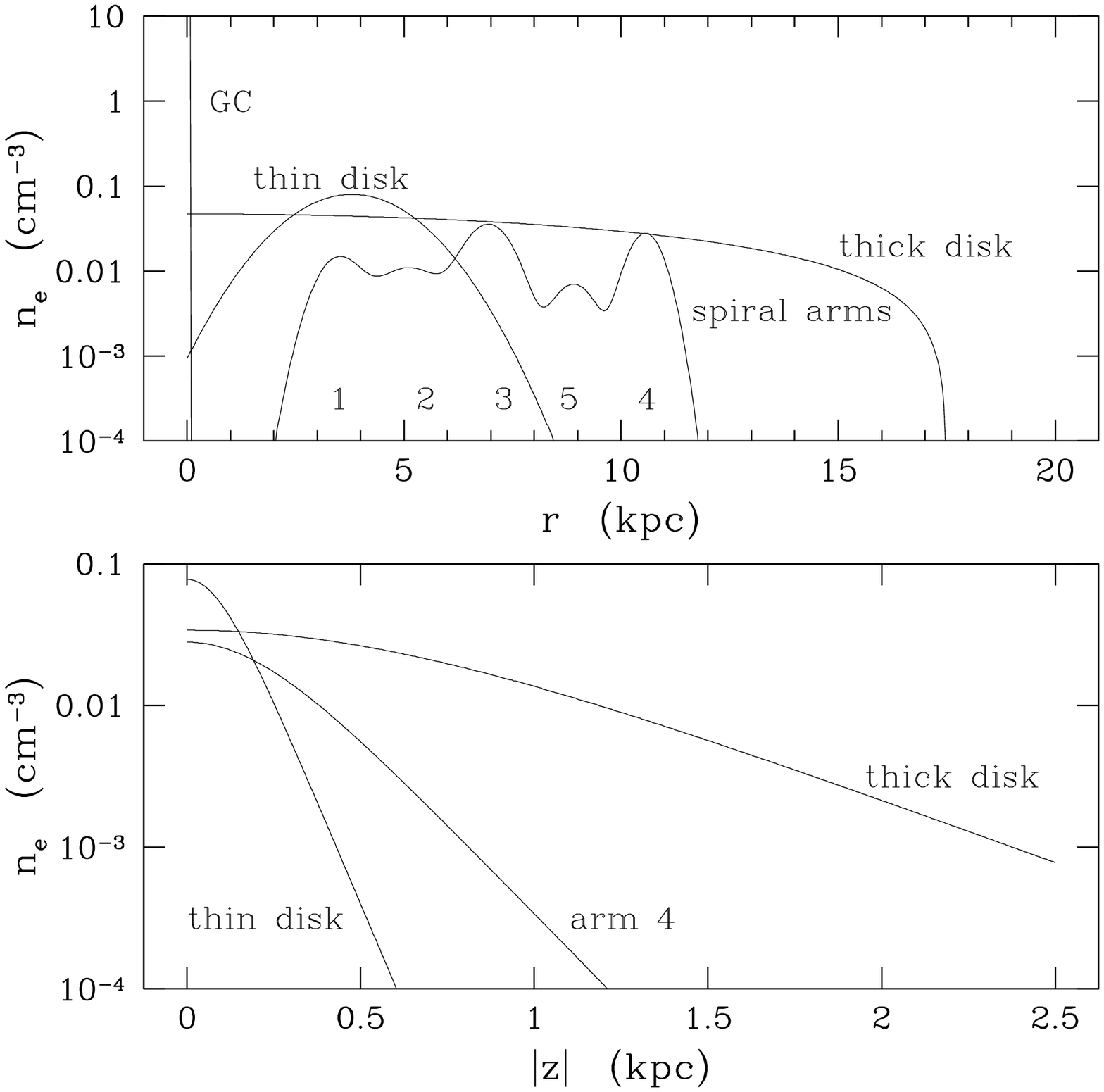}
\figcaption{
\label{fig:modelcuts}
\textit{Top}: Electron density plotted against Galactocentric radius
in the direction from the Galactic center through the Sun for the
various large-scale components.  
\textit{Bottom}: Electron density plotted against $\vert z\vert $.
For the inner Galaxy (thin disk) component, the profile is for 
$r = 3.5$ kpc, the peak of the annular component.
For the thick disk component, the cut is at $r=R_{\odot}$.
The spiral arm cut is at $(x,y) = 0,10.6$ kpc.
}

\subsubsection{Outer, Thick Disk Component} \label{sec:thick} \label{sec:outer} 
The outer, thick disk component is responsible for the DMs of
globular cluster pulsars and the low-frequency diameters of
high-latitude extragalactic sources (e.g., as inferred from
interplanetary scintillation measurements).  In TC93 this component
was determined to have a scale height of roughly 1~kpc with a
Galactocentric radial scale length of roughly 20~kpc.  However, the
data available for TC93 did not allow a firm
constraint on the Galactocentric scale length; scale lengths as large
as 50~kpc were also allowed by the data.
Through measurements of scattering of extragalactic sources toward
the Galactic anticenter,  Lazio \& Cordes (1998a,b) inferred a scale
length $\sim 20$ kpc and a functional form that truncates at this
scale.  Alternatives are discussed in Paper II.

\subsubsection{Inner, Thin Disk Component} \label{sec:thin} \label{sec:inner}

An inner Galaxy component ($n_2$) consisted of a Gaussian annulus in
TC93.  Data available to TC93 could not distinguish a filled Gaussian
form in Galactocentric radius from an annular form, but the latter was
chosen for consistency with the molecular ring seen in CO (e.g.,
\cite{dameetal87}.  We adopt the same form in our model; in Paper II
we discuss alternatives to the annular form.   

\subsubsection{Spiral Arms}

Is spiral arm structure required by pulsar dispersion measures 
and distance constraints? 
TC93 argue that it is by referring to the asymmetry of the
distribution of \DM\ vs.\ Galactic longitude (c.f.\ their Figure 2). 
The same asymmetry appears in the larger sample now available (Paper II).
A direct demonstration can be made by calculating the \DM\ deficit for
individual pulsars 
for various electron-density models, defined as the difference
between the model \DM\ integrated to infinite distance and the pulsar \DM: 
$\Delta\DM = \DM - \DM_{\infty}({\rm model})$.
In Figure~\ref{fig:gbc01lb} we show $\Delta\DM$ plotted against 
Galactic coordinates  for the axisymmetric model of GBC01.  
There is a 
large number of \DM\ deficits, both along the Carina-Sagittarius
arm ($\ell \sim -65^{\circ}\pm10^{\circ}$)
and extending (at low latitudes) continuously to $\ell = +60^{\circ}$.
This broad longitude range encompasses all of the spiral arms interior
to the solar circle and the molecular ring.    
GBC01 identify the need for spiral structure in 
similar directions by comparing predicted and actual DMs for the 
much smaller data set they considered.    The large number of deficits
we identify with their model (183 out of 1143 objects)
is a much stronger signal for spiral structure.  Spiral arm structure
appears to be mandatory in any electron density model that aspires to
realism. 

The TC93 model, though containing spiral structure, also has insufficient
\DM\ (Figure~\ref{fig:tc93lb}) 
along the Carina-Sagittarius arm and at high latitudes.    The low latitude
deficits are removed in the new model by redefining and refitting the
spiral arms while the high latitude deficits are removed by refitting
the outer, thick-disk component and the spiral arm scale heights.

\medskip
\epsfxsize=8truecm
\epsfbox{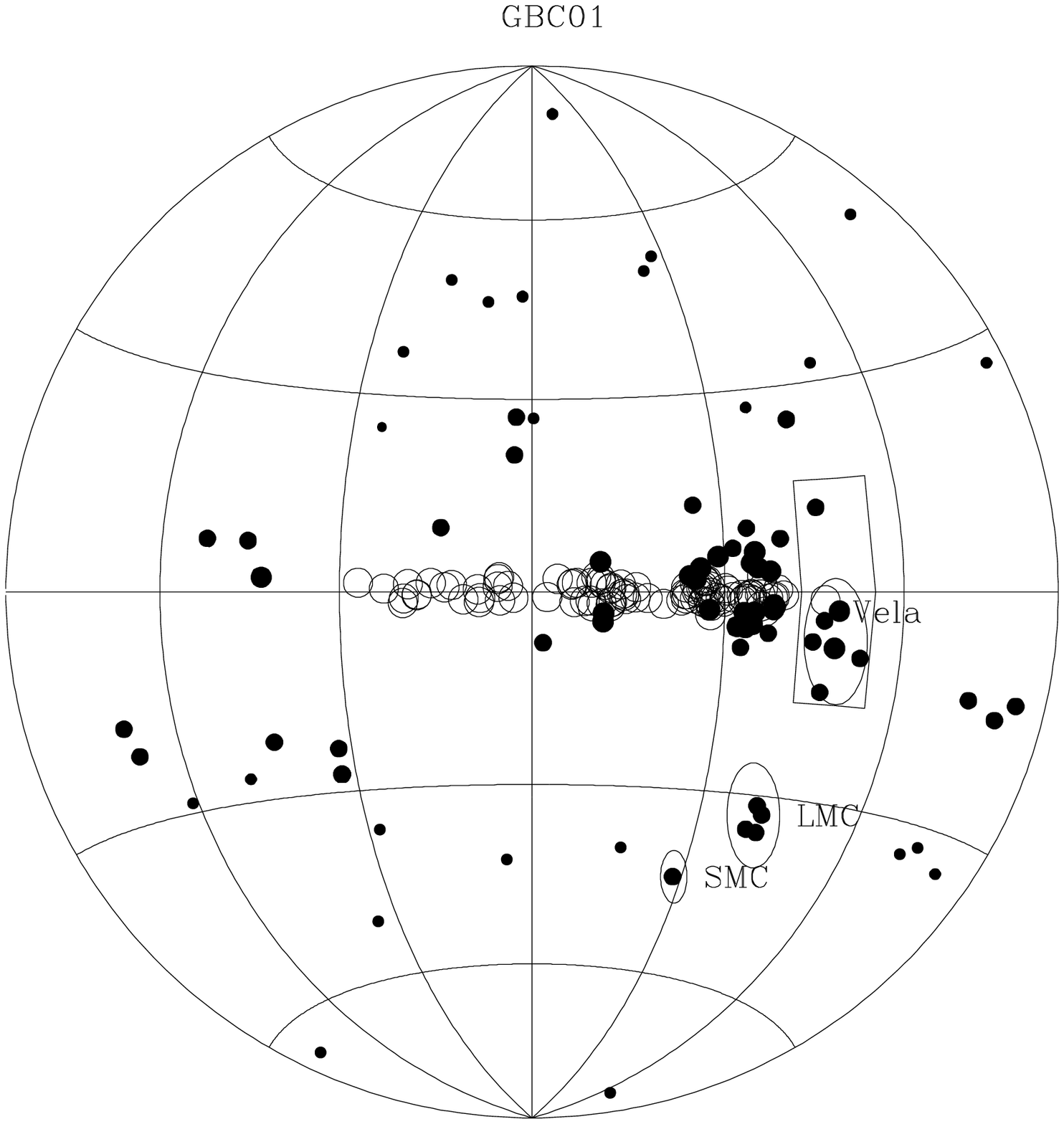}
\figcaption{
\label{fig:gbc01lb}
Plot of \DM\ deficit,
$\Delta \DM = \DM - \DM_{\infty}$(GBC01)  against Galactic coordinates,
where $\DM_{\infty}$(GBC01) is the maximum \DM\ obtained by 
integrating the axisymmetric model of  Gomez, Benjamin \& Cox (2001; GBC01) 
to infinite distance.
Longitudes $0^{\circ}$- $180^{\circ}$ are on the left.
Only positive residuals are shown, yielding 183 objects
out of the 1143 pulsars in the
combined Princeton and (public) Parkes Multibeam samples.
The filled circles (going from smallest to biggest) represent 
$\Delta\DM<50$,
$50<\Delta\DM\le 200$, 
and
$200<\Delta\DM\le 400$ pc cm$^{-3}$. 
The open circles are for $\Delta\DM>400$ pc cm$^{-3}$.
Ellipses designate pulsars in the Large and Small Magellanic
Clouds and the Vela supernova remnant region.  The rectangular
region designates objects affected by the Gum Nebula. 
The GBC01 model is clearly deficient in a number of directions,
including the cluster of objects near 
$(\ell, b) = -65^{\circ}\pm10^{\circ}, 0^{\circ}$, 
which indicates insufficient
column density along the Carina-Sagittarius spiral arm.    
A large
number of objects have $\Delta\DM > 0$ from longitudes
$-65^{\circ}$ to +60$^{\circ}$, indicating the need for
spiral structure in their model like that in 
TC93  and in the present model.
}

\medskip
\epsfxsize=8truecm
\epsfbox{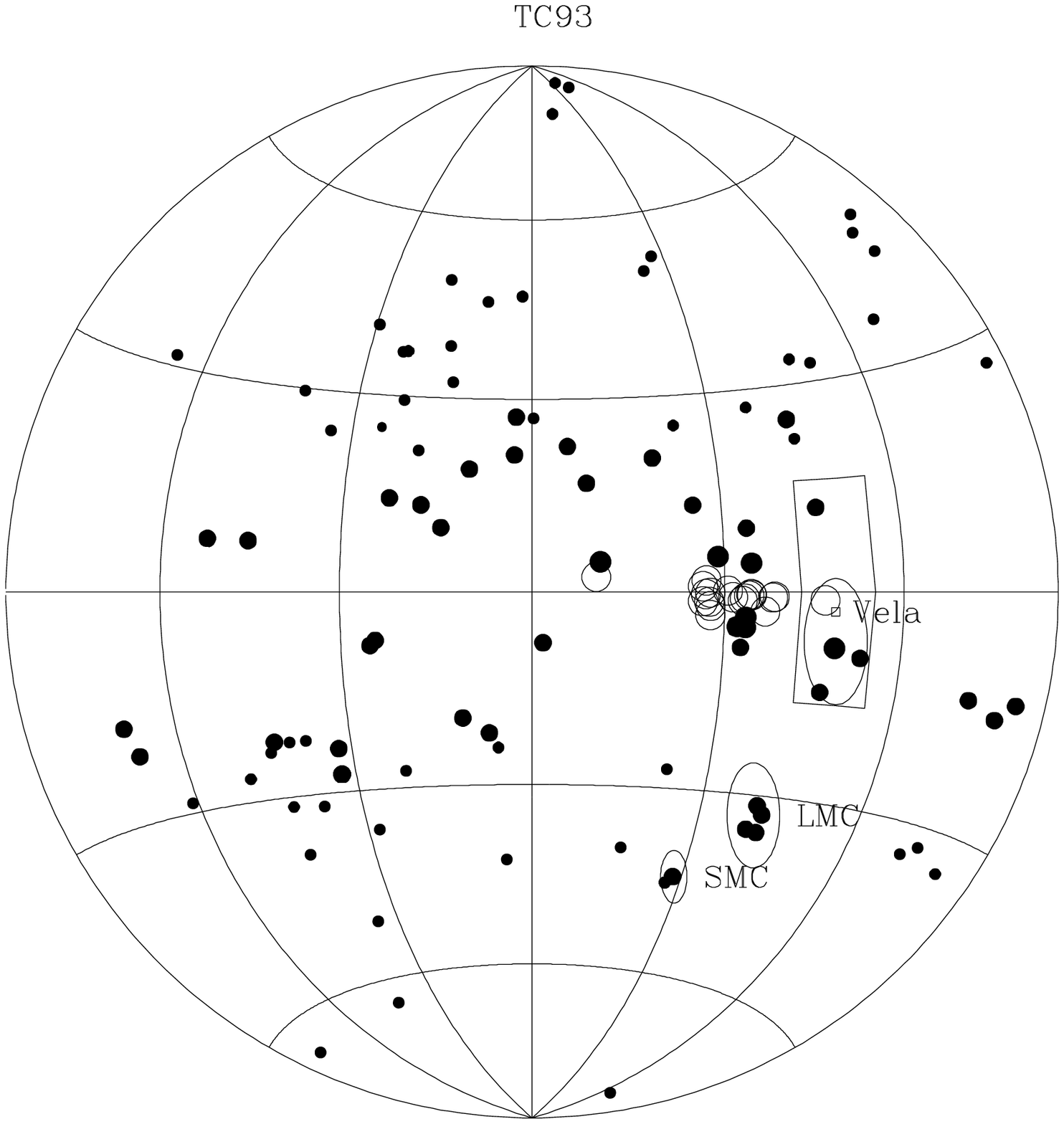}
\figcaption{
\label{fig:tc93lb}
Plot of \DM\ deficit,
$\Delta \DM = \DM - \DM_{\infty}$(TC93)  against Galactic coordinates,
where $\DM_{\infty}$(TC93) is the maximum \DM\ obtained by 
integrating the TC93 model  to infinite distance.
The format is the same as in Figure~\ref{fig:gbc01lb}.
The TC93 model is clearly deficient in a number of directions, 134 out of
1143 pulsars having values of \DM\ too large to be accounted for,
including the cluster of objects near 
$(\ell, b) = -65^{\circ}\pm10^{\circ}, 0^{\circ}$, 
which indicates insufficient
column density along the Carina-Sagittarius spiral arm.    
The new model corrects these deficiencies. 
}

\medskip
The spiral arm  centroids in the new model are defined as logarithmic spirals
with perturbed locations similar to those used in TC93. 
Centroid parameters are similar to those of Wainscoat \etal\ (1992).
The spiral structure is similar to that proposed by Ghosh \& Rao (1992)
and also described by Vall\'ee (1995, 2002).   We have maintained usage
of a Sun-Galactic Center distance of 8.5 kpc, the IAU recommended
value,  although recent work favors a smaller distance,
$\sim 7.1\pm0.4$ kpc
(e.g., Reid 1993; Olling \& Merrefield 1998)\footnote{In the next version
of the model, we will explore usage of a smaller Sun-GC  distance.}.
Details are given in Paper II.  The new model includes a local (Orion-Cygnus)
spiral arm.   Also, each arm has an individual centroid electron density,
width, scale height and $F$ parameter. 
Figure~\ref{fig:armdefs} shows the locations of spiral arms
as defined in TC93 and as modified by us.

We emphasize that 
the spiral arm components in our model, like
those in TC93, are modeled as overdense regions.  Astrophysically, 
however, the enhanced star formation in spiral arms will produce
underdensities  as well as overdensities, as has been demonstrated
by the identification of supershells.  Though we have
adopted spiral arms as overdense regions, to account for \DM\
and \SM\ we have had to introduce  ellipsoidal underdense perturbations
(``voids'') in particular directions (\S\ref{sec:voids}).

\medskip
\epsfxsize=8truecm
\epsfbox{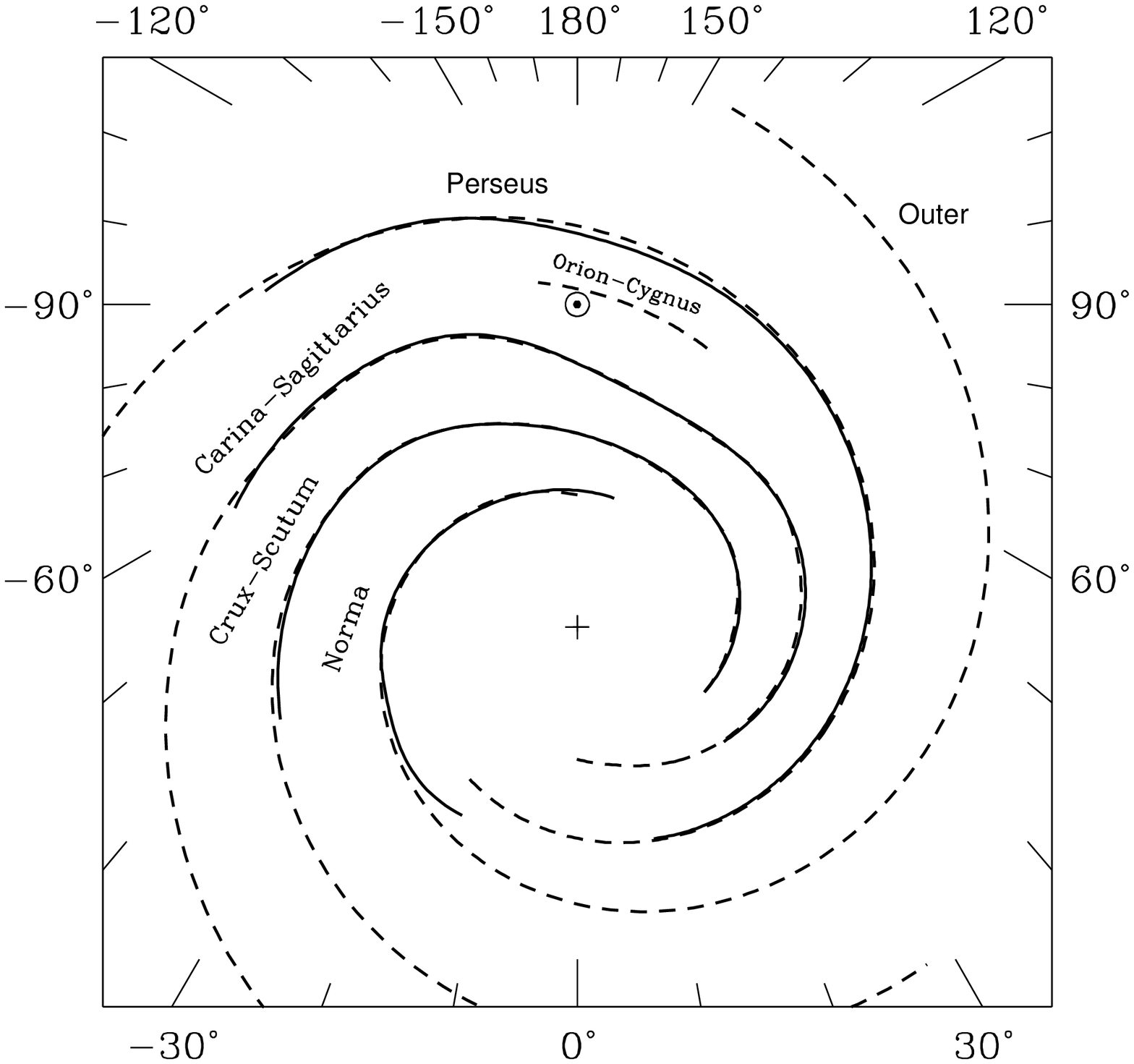}                
\figcaption{
\label{fig:armdefs}          
Solid lines: spiral model of the Galaxy used in TC93,
defined according to work by  Georgelin \&
Georgelin (1976), modified as in TC93.
Dashed lines:  a four-arm logarithmic spiral model combined
with a local (to the Sun) arm using
parameters from Table 1 of Wainscoat \etal\ (1992),
but modified so that the arms match some of the features of
the arms defined in TC93.
The names of the spiral arms, as in the astronomical literature,
are given.
A $+$ sign marks the Galactic center and the Sun is
denoted by $\odot$. 
}

\subsection {Local ISM (LISM)}\label{sec:lism}


We model the local ISM in accord with \DM\ and \SM\ measurements of
nearby pulsars combined with parallax measurements  and guided by 
\halpha\ observations that provide estimates of \EM.  
Observations and analysis by Heiles (1998),
Toscano \etal\ (1999 and references therein), Snowden \etal\ (1998),
Ma\'iz-Apell\'iz (2001) and others (see Appendix A)
suggest the presence of four regions of low density near the Sun:
(1) a local hot bubble (LHB) centered on the Sun's location;
(2) the Loop I component (North Polar Spur) that is long known because
	of its prominence in nonthermal continuum maps;
(3) a local superbubble (LSB) in the third quadrant; and
(4) a low density region (LDR) in the first quadrant.
Additional features have been identified by Heiles (1998) but the available
lines of sight to pulsars appear to not require their inclusion in our
model. 
Bhat \etal\ (1999) explicitly fitted for parameters of the LHB
using pulsar measurements and a model having a low-density structure 
surrounded by a  shell of material that produces excess scattering.
Some of the parallax distances used by Toscano \etal\ (1999)
have been revised, in some cases substantially, implying lower densities
in the third quadrant than they inferred. 

Appendix A describes the mathematical models used for the four regions.  
Table \ref{tab:lism} lists the parameters of the local ISM model
and their values based on the fitting we describe in Paper II.

\medskip
\epsfxsize=8truecm
\epsfbox{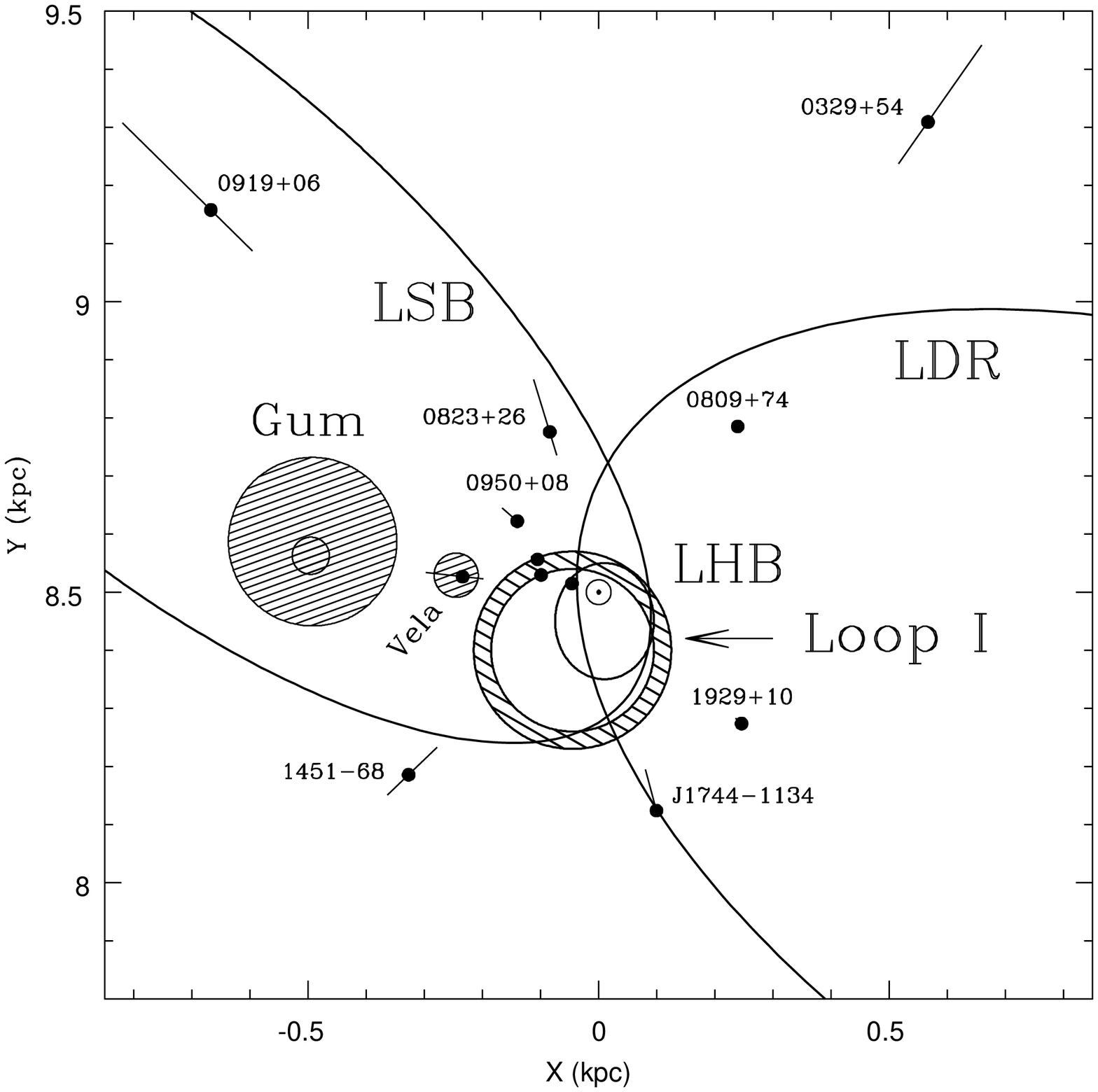}
\figcaption{
\label{fig:lism}
Projection onto the Galactic (X-Y) 
plane of the four local interstellar medium components,
LHB, LSB, LDR and Loop I along with the clumps defining
the Gum Nebula and the Vela supernova remnant.  
The LSB and LDR are the large ellipsoids, Loop I is the annular
hatched region and the LHB is the small ellipsoid containing
the Sun ($\odot$). 
Filled circles
show the DM-predicted locations using NE2001 of those pulsars
having parallax measurements. 
The lines plotted through each point (not all are 
visible) represent the allowed distance ranges from the parallax measurements. 
The three unlabeled points closest to the Sun are pulsars
B1237+25, J0437$-$4715, and B1133+16 in order of increasing 
projected distance from the Sun.
}
\medskip

\subsection{Galactic Center (GC) Component}\label{sec:gc}

A new component is the region in the Galactic center that is responsible
for scattering of \sgra\ and OH masers 
(van Langevelde \etal\ 1992; Rogers \etal\ 1994; Frail \etal\ 1994). 
Typical diameters, scaled to~1~GHz, are approximately 1\arcsec, roughly 10 times
greater than that predicted by TC93, even with the general enhancement
of scattering toward the inner Galaxy in that model.
The model of the  region is based on work by Lazio \& Cordes~(1998c,d) 
who used angular sizes of extragalactic sources
viewed through Galactic center to localize and model the region.  

We use a slightly altered version of the axisymmetric model 
of Lazio \& Cordes~(1998d).
We use  a scale height $\HGC\approx 0.026$ kpc and slightly smaller
radial scale, $\RGC\approx 0.145$ kpc.  In addition we have offset
the center of the distribution by 
$(x_{\rm GC}, y_{\rm GC}, z_{\rm GC}) = (-0.01, 0, -0.02)$ kpc.
The offset ellipsoid exponential is truncated to zero for arguments 
smaller than $-1$.
A value $F_{\rm GC} \sim 6\times 10^4$ 
produces SM values needed to account for the scattering diameters 
of \sgra\ and the OH masers.

\subsection{Gum Nebula and Vela Supernova Remnant} \label{sec:gum}

In TC93 a large region
was included which perturbed the dispersion measures of pulsars viewed 
through the Gum Nebula but did not influence the scattering. 
Since the writing of TC93,  investigations of the Vela pulsar and other
pulsars indicate that scattering is large within a region
of at least 16 degrees diameter centered roughly on the direction
of the Vela pulsar  (Mitra \& Ramachandran 2001).   We have modeled
the Gum/Vela region based on the scattering measurements and also
on the fact that enhanced, local mean electron density is required
to account for the dispersion measures and/or distances of five objects.

We model the direction toward the Gum Nebula
with an overlapping pair of spherical regions
describing the Gum Nebula itself and another for 
the immediate vicinity of
the Vela pulsar.  These are included in the list of ``clumps'' that 
enhance the electron density and $F$ (see below).

\subsection{Regions of Intense Scattering (``Clumps'')}

In addition to regions that perturb the scattering of
the Vela pulsar and other pulsars near it and the Galactic center, other
regions of intense scattering must exist to account for the large
angular diameters and/or pulse broadening seen toward a number of
Galactic and extragalactic sources.  
We define ``clumps'' as regions of enhanced $n_e$ or $F$ or both
and identify them by iterating with preliminary fits to the smooth
components of the electron-density model (e.g., the thin and thick disk
components and the spiral arms).  Clumps are effectively
manifestations of the Galaxy's mesoscale structure not modeled
adequately by the assumed large-scale components.

We model clumps with
thickness $\Delta s \ll D$, and we use parameters $\nec, \Fc$ and $\dc$
(electron density inside the clump, 
fluctuation parameter, and distance from
Earth).  The implied increments in \DM\ and \SM\ are
\be
\DM_c &=& \nec\Delta s \nonumber \\
\SM_c &=& \csm~\Fc~\nec^2 \Delta s = 10^{-5.55} \DM_c^2 F_c / \Delta s_{kpc},
\label{eq:smclump}
\ee
where the last equality holds for \DM\ and \SM\ in standard units.
Equation~\ref{eq:smclump} implies that relatively modest contributions
to \DM\ can produce large changes in \SM\ if the clump is small and 
fluctuation parameter large.  For example, 
$\DM_c = 10$ pc cm$^{-3}$, $F_c = 1$ and $\Delta s = 0.02$ kpc
yield $\SM_c = 0.014$ kpc m$^{-20/3}$.   For a pulsar 1 kpc away,
the clump would perturb \DM\ and  the resultant
distance estimate by perhaps 30\% while \SM\ would increase by about 
a factor of 100.   Thus pulsars which have anomalous scattering
relative to the smooth model can, in many cases,
 be well modeled with only small perturbations to the \DM\ predictions
of the model.  This also means that the model is expected to
be much better for distance and \DM\ estimation than for
scattering predictions.     

Table~\ref{tab:clumpsX} lists clumps needed to account for the scattering
toward specific AGNs that have enhanced scattering;  most are in the
Cygnus region.
Table~\ref{tab:clumpsG} lists clumps needed to account for the scattering
of non-pulsar Galactic sources; all are OH masers except for the
Galactic continuum source \objectname[]{Cyg X-3}.
For these non-pulsar sources, only the scattering is measured and so
the clump perturbation, $\SMc$, is the most robustly determined clump
parameter.   We adopt reasonable --- but non-unique --- values for
the clump distance, radius, and fluctuation parameter
which imply a value for the clump's $\DM_c$, which we also tabulate.  

For pulsar lines of sight, on the other hand, the clump's
contributions to both \DM\ and \SM\ are constrained if a scattering
measurement exists.
Included in Table~\ref{tab:clumpsP.short}
are those clumps that have $\DMc >  20$ pc cm$^{-3}$ {\it or\/}
$\SMc > 1$ kpc m$^{-20/3}$.  Operationally, the model includes additional,
weaker clumps that bring the scattering and distances into accord with
observations.  The software implementation of the model (see below)
includes these weaker clumps as well as those listed in
Tables~\ref{tab:clumpsX}--\ref{tab:clumpsP.short}.

In most cases our fitting procedure does not provide unambiguous
distance estimates for the clumps.  The distances listed in
Tables~\ref{tab:clumpsX}--\ref{tab:clumpsP.short} are ``plausible''
distance estimates for clumps, locating them, for instance, in or near
spiral arms or in specific HII complexes that contain molecular masers.

\subsection{Regions of Low Density (``Voids'')}\label{sec:voids}

We found that some pulsar distance constraints could not be satisfied
using the previously defined structures without recourse to placement
of a low-density region along the line of sight.  We call these
regions ``voids'' although they simply represent, typically,
regions with lower-than-ambient density.   By necessity, they take precedence
over all other components (except clumps), which we effect by usage of a void
weight parameter, $\wvoids = 0,1$, that operates similarly to
$\wlism$.   The mathematical form is given in Table~\ref{tab:ne2001}.
We use elliptical gaussian functions with semi-major and semi-minor
axes $a, b, c$ and a rotation angle $\theta_z$ about the
 $z$ axis.  Table~\ref{tab:voids} lists the properties of
all voids used in the current realization of the model. 




\section{Model Performance}\label{sec:performance}


The quality of the model can be evaluated on the basis of
how well it estimates the distances and scattering
of pulsars with appropriate measurements.

Figure \ref{fig:lism} shows nearby pulsars with
parallax distance ranges plotted as lines projected onto
the Galactic plane.   Extrapolations of these lines 
intersect the Sun's location.   The filled circles
indicate the distance estimated with NE2001.
As can be seen, distance estimates for 
most of the nearby pulsars are acceptable.   A uniform medium
or one with only large-scale structure (i.e. one without 
the Gum Nebula, Vela supernova remnant, and LISM components)
would do much more poorly.

Figure \ref{fig:gplane_ddm} shows a projection onto the Galactic
plane of distance ranges and model estimates for those pulsars
having independent distance constraints and Galactic latitudes,
$\vert b \vert < 5^{\circ}$.  Only six objects
have model distances that fall outside the distance range.
As commented upon in Paper II 
two objects have
questionable lower distance bounds from \ion{H}{1} absorption.  
The other four lines of sight are toward pulsars in low-latitude
globular clusters toward the Galactic center.
Also shown in the figure
along the perimeter (as filled squares) are objects for which
the TC93 model could provide only a lower distance bound.
These occur in the directions along the Carina-Sagittarius spiral
arm, through the Gum Nebula, and toward the inner Galaxy near
$\ell \approx -20^{\circ}$.   The new model accounts for the DMs of
these objects by having larger electron densities along the 
relevant lines of sight.

Figure \ref{fig:dhat.vs.d} shows the DM-calculated
distance plotted against independent distance
constraints, $[\dl, \du]$.   Model estimates are
bracketed by $[\dl,\du]$ in the large majority of
cases (90 of 120 total, though not all are plotted and
some of the 30 non-bracketed cases are multiple pulsars in
the same globular cluster).   For most cluster pulsars, the
distance is underestimated because the cluster is well
beyond the electron layer of the Galaxy.   

Figure \ref{fig:rhist} quantifies the model's ability to account for
the scattering of known objects  as a histogram of the quantity
$\rscatt$ = (predicted scattering)/(measured scattering).
The large peak centered on $\rscatt=1$ indicates that most lines of sight
are well modeled, though some outliers remain.
The outliers are largely globular cluster pulsars whose distances
are poorly estimated by the model 
 and a few objects whose lines of sight are 
dominated by the local \hbox{ISM}.  It should be emphasized that many of
the small scale features in the present model were introduced to
yield a good prediction of the scattering, so the histogram
is largely a manifestation of our modeling approach.   The predictive
aspect of the model will be tested when it is applied to
objects on which new scattering measurements are obtained. 

\smallskip
\epsfxsize=8truecm
\epsfbox{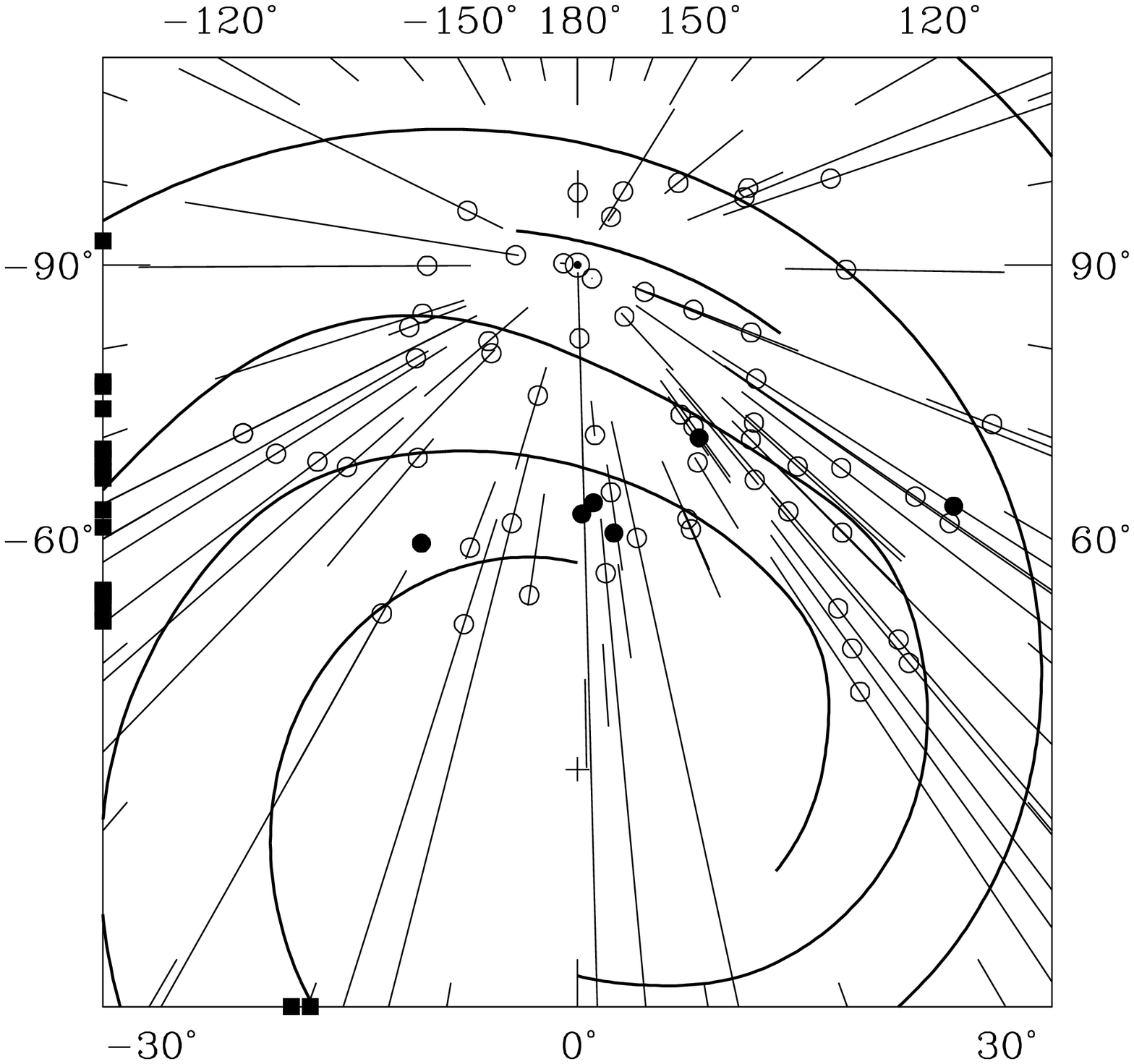}
\figcaption{
\label{fig:gplane_ddm}
Distances calculated from model NE2001 projected onto the
Galactic plane for objects with $\vert b \vert < 5^{\circ}$.   
In the interior of the box,
plotted lines indicate independent distance ranges, 
$[\dl, \du]$,  open circles denote ``good'' fits 
(i.e. model distances $\hat D$ that are within the empirical
range) while the six filled circles designate 
pulsars where the model distance underestimates the minimum
empirical distance.  Two of these objects
(B1930+22, $\ell = 57.4^{\circ}$, $b = 1.6^{\circ}$;  
 J1602$-$5100, $\ell = -29.3^{\circ}$, $b = 1.3^{\circ}$)
 have uncertain lower 
distance bounds from HI absorption while the other four
are globular cluster pulsars at low latitudes toward the 
Galactic center ($0^{\circ}\lesssim \ell \lesssim 40^{\circ}$). 
On the perimeter of the box, filled squares denote 
pulsars with $\vert b\vert \le 5^{\circ}$ for which the TC93
model fails to yield distance estimates  
because their dispersion measures exceed those 
of the model: $\DM > \DM_{\infty}({\rm TC93})$.  These points
are at or near tangents to the spiral arms in the
fourth quadrant ($-83^{\circ} \le \ell \le -67^{\circ}$)
or are along the line of sight through the Gum Nebula.
The new model corrects these defects and 
has $\DM({\rm NE2001}) > \DM$ for all objects except those
in the Magellanic clouds.
}

\medskip
\epsfxsize=8truecm
\epsfbox{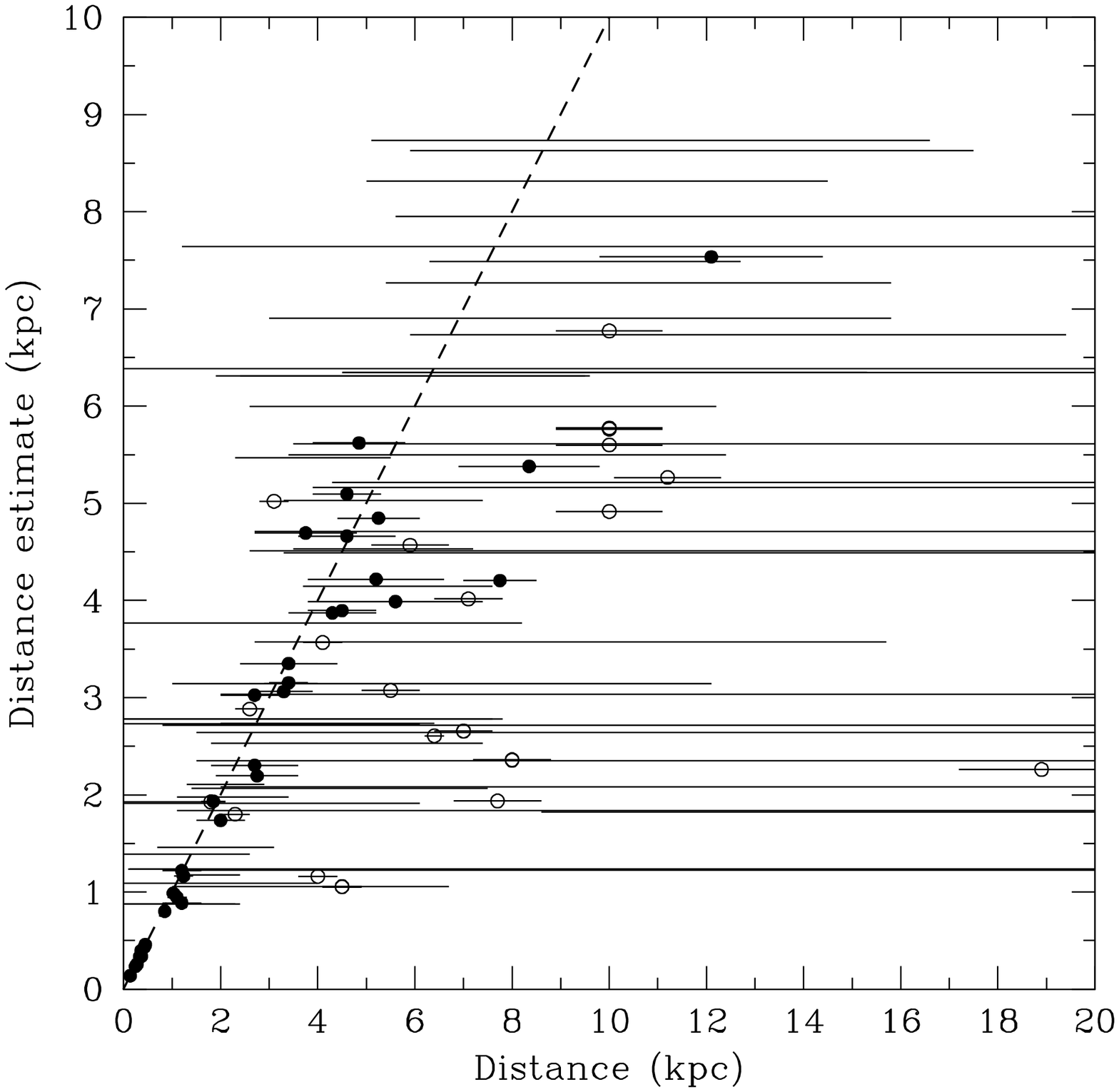}
\figcaption{
\label{fig:dhat.vs.d}
Distances calculated from the model plotted against independent 
empirical ranges, $[\dl, \du]$, shown as horizontal lines.    
A filled circle is plotted
at $(\dl+\du)/2)$ when the range is relatively small,
$\du/\dl < 2$.  
For globular cluster pulsars, an open circle is plotted.
}

\medskip
\epsfxsize=8truecm
\epsfbox{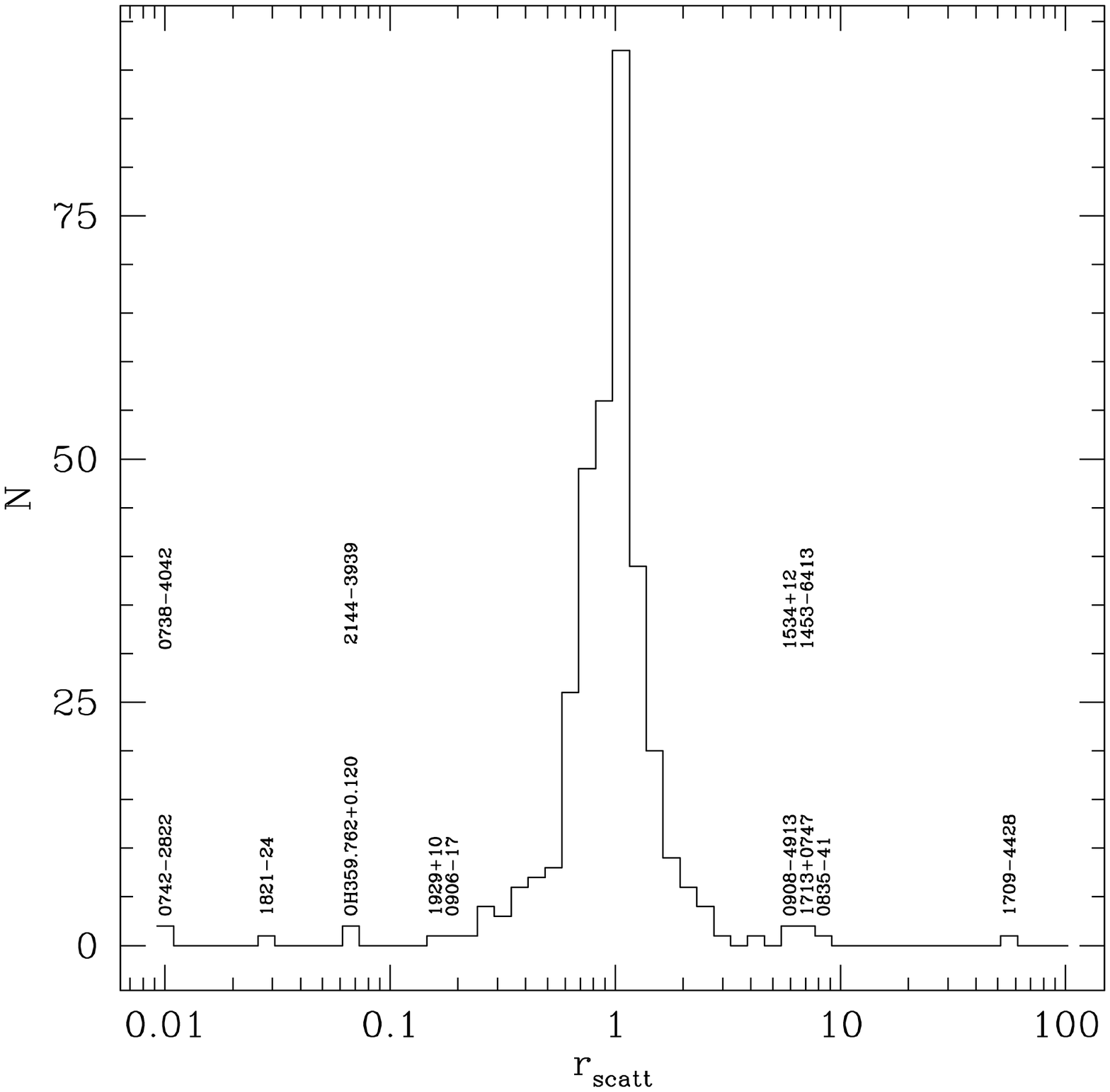}
\figcaption{
\label{fig:rhist}
Histogram of the scattering prediction ratio, $\rscatt$,
defined as $\rscatt = $ predicted / actual value of scattering
observable for angular and pulse broadening and
$\rscatt = $ actual / predicted for scintillation bandwidths. 
Thus $\rscatt>1$ ($\rscatt<1$) means scattering is 
overpredicted (underpredicted). 
}

\subsection{Asymptotic Values for DM}


Figure \ref{fig:dmvsl} shows asymptotic values for \DM\ in our
model, $\DM_{\infty}$, plotted against Galactic longitude.
In the inner Galaxy, the asymptotic values exceed the highest
known \DM\ (1209 pc cm$^{-3}$) by a substantial factor.  The lack of
pulsars with such large values of DM is consistent with various
selection effects, as we discuss in Paper II.  
For instance, the channel bandwidths used in the
Parkes survey imply dispersion smearing of order $15$ ms for $\DM =
2000$ pc cm$^{-3}$, comparable to the pulse width of many pulsars.
More importantly, the pulse broadening from scattering will exceed 1 s
for many objects.  Combined with inverse square law effects, one would
not expect to find pulsars with much larger values of DM to be present in
the existing surveys.

\medskip
\epsfxsize=8truecm
\epsfbox{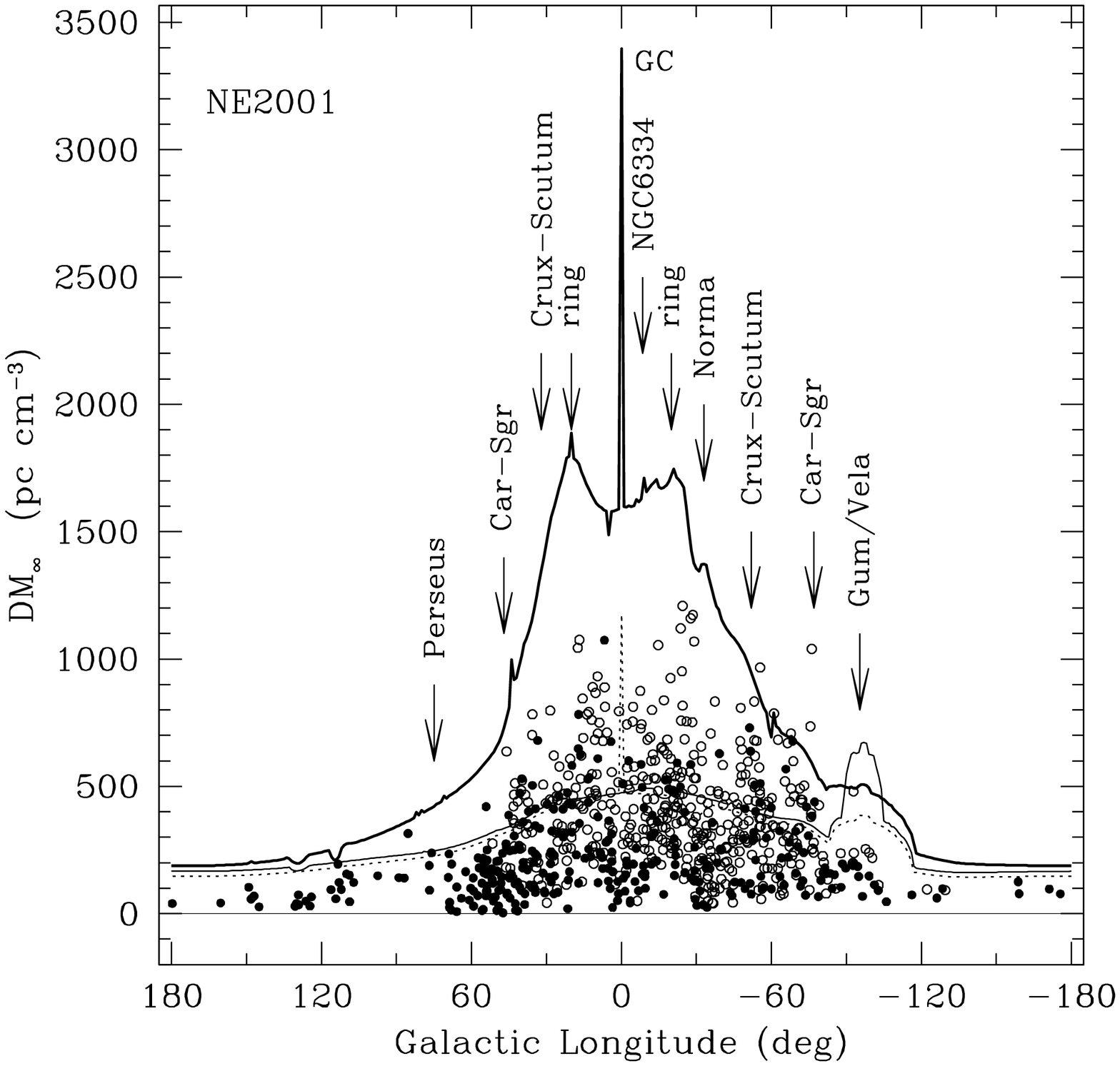}
\figcaption{
\label{fig:dmvsl}
Plot of $\DM_{\infty}(\ell, b)$, 
the maximum \DM\ obtained by integrating the NE2001 model.
Heavy solid line: $b = 0^{\circ}$.
Light solid line: $b = -5^{\circ}$.
Dotted line: $b = +5^{\circ}$ (offset by -20 pc cm$^{-3}$ in the
vertical direction for clarity).
Plotted points are pulsars with $\vert b \vert < 5^{\circ}$.
Filled circles: 
pulsars from the Princeton catalog.
Open circles: 
pulsars from the public Parkes Multibeam catalog.
Labelled features include tangents to the spiral arms and
maxima associated with the ring component, the Gum Nebula and
Vela supernova remnant, and the \ion{H}{2} complex \protect\objectname[NGC]{NGC~6334}.
The gaps near $\ell = +15^{\circ}$ and $-20^{\circ}$ 
and $\DM \lesssim 200$ pc cm$^{-3}$ are real,
appearing in both samples, and signify either the real absence of pulsars
in the corresponding volume or  the presence of large electron
densities in these directions fairly close to the solar system.
The absence of pulsars above $\sim 1200$ pc cm$^{-3}$ is due to selection against
such objects by pulse broadening from dispersion smearing and scattering.
}
\medskip

\subsection{Distance Errors to be Expected}

An ionized feature not included in our model will perturb
the dispersion measure toward an individual object by an amount
$\Delta\DM$.
From the definition of \DM, it follows that the
consequent distance error, if small,  will be
\be
\Delta D = \frac{\Delta\DM}{n_e(\xvec(D))}.
\label{eq:derror}
\ee
Thus, for given $\Delta\DM$,
the distance error is larger for a pulsar in a region
of small electron density compared to one in a dense region.
(Of course, the perturbation can be large enough to render
a very large distance error not describable by Eq.~\ref{eq:derror}.)
A particularly important case is for an object that is more than
one scale height above the Galactic disk.   Using a simple
plane-parallel model as an example, the distance error is
\be
\Delta D = \frac{\Delta\DM}{{n_e}_0 (1 - \DM/\DMinfty)},
\ee
where $\DMinfty = {n_e}_0 H/\sin\vert b\vert$ is the maximum
\DM\ of the model with midplane density ${n_e}_0$ and
scale height $H$  toward a direction at Galactic latitude
$b$.   For objects with \DM\ near the asymptotic value,
distance errors are much larger than for objects within one
scale height of the plane.    This is true for both
positive and negative perturbations from density enhancements
and voids, respectively.  For large positive perturbations of
\DM\ the measured \DM\ can exceed $\DMinfty$ (or its analog in
our multicomponent model).  In this case, the model can yield
only a lower bound on the distance.

For NE2001, we have explicitly constructed the model so that nearly 
all known pulsars have $\DM < \DMinfty$.    This feature may introduce
bias in the model because there very well could be pulsars at large distances
with DMs that are small because an unmodeled  void is present 
along the line of sight.   We estimate that the number of lines of sight
for which this would be true is a small fraction of the total pulsar
sample.

In Figure \ref{fig:derrors} 
we show the distance errors that ensue when we perturb
\DM\ by amounts corresponding to \ion{H}{2} regions of different kinds.
For reference, we expect perturbations in \DM\ with the following
amplitudes
(e.g. Prentice \& ter Haar 1969; Bronfman \etal\ 2000): 
\begin{enumerate}
\itemsep -2pt
\item
Str\"omgren sphere around an O5 star:
$\Delta DM \approx 75$ pc cm$^{-3}$. 
\item
Str\"omgren sphere around an O9/B1 star:
$\Delta DM \approx 3$ to 10 pc cm$^{-3}$. 
\item 
OB association: $\Delta\DM \approx 100$ to 200 pc cm$^{-3}$.
\item
Ultra-compact \ion{H}{2} region:
$\Delta DM \approx 300$ to 500 pc cm$^{-3}$. 
Through a simulation of pulsar and UCHII regions born in spiral arms
and the molecular ring, we find that 
one expects only about one pulsar out of the observed population to
intersect an UCHII region.
\end{enumerate}

\medskip
\epsfxsize=8truecm
\epsfbox{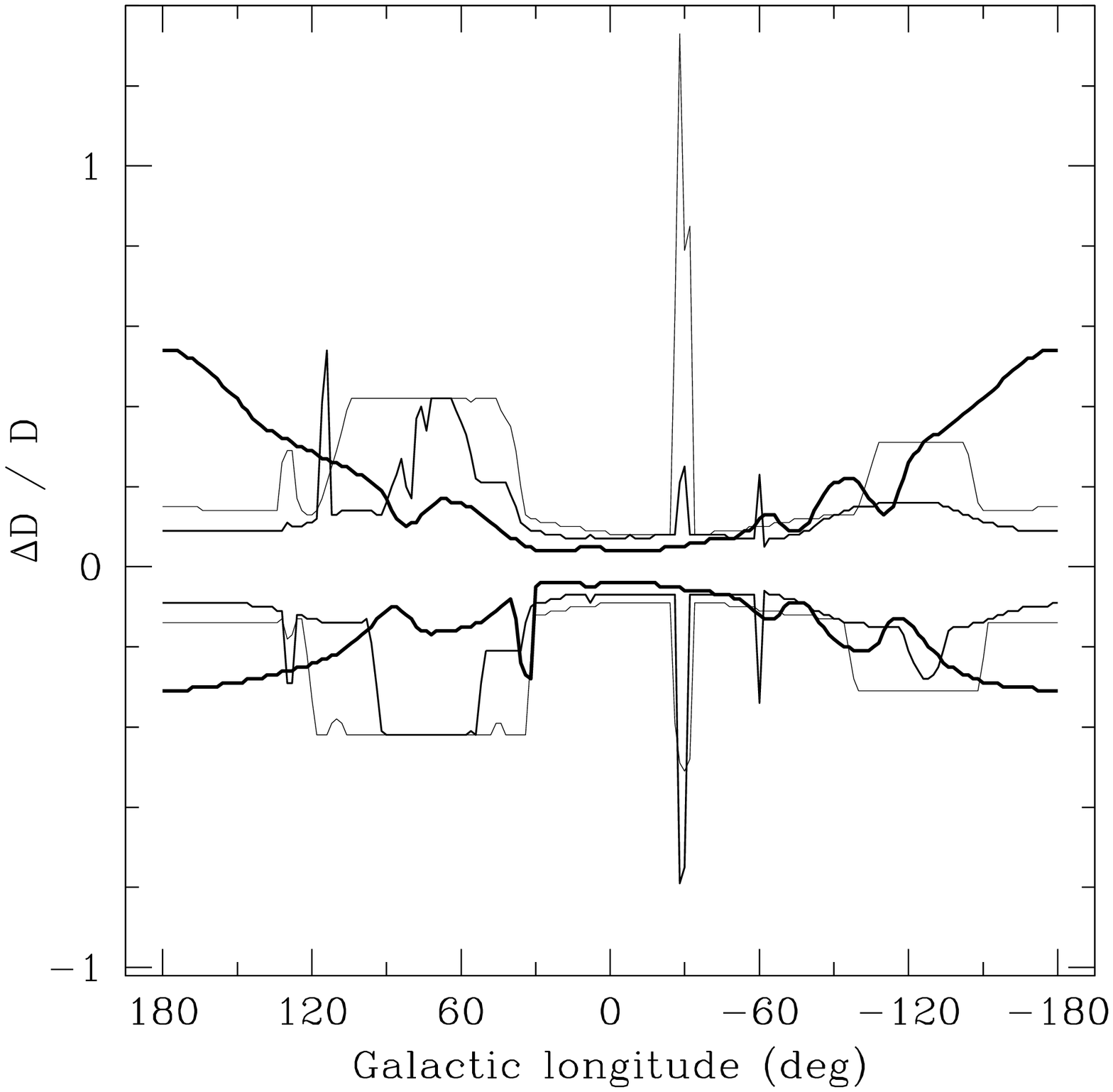}
\figcaption{
\label{fig:derrors}
Plot of fractional distance error, $\Delta D / D$, against
Galactic longitude where $\Delta D = \hat D - D$.   
The curves are calculated for
perturbations to \DM\ of $\pm\Delta\DM$ which increase or decrease
the model distance, $\hat D$, from the actual distance.
Thickest line: $D = 5$ kpc and $\Delta\DM = 30$ pc cm$^{-3}$.
Medium line: $D = 2$ kpc and $\Delta\DM = 10$ pc cm$^{-3}$. 
Thinnest line: $D = 1$ kpc and $\Delta\DM = 5$ pc cm$^{-3}$. 
}

\subsection{Comparison with TC93 Distances}\label{sec:compare}

Figure \ref{fig:d2001vsdtc93} compares distance estimates from
NE2001 with those from TC93.  The general trend is that
the TC93 distances are larger, in some cases significantly so,
than NE2001 distances;  there are also cases where
NE2001 provides smaller distances.
The generally larger TC93 distances are 
consistent with our other diagnostics,
which signify that TC93 provides too few electrons along
many lines of sight.   As designated in the figure,
TC93 fails to give a distance estimate for
134 lines of sight. The smaller distances from NE2001
will influence estimates of pulsar space velocities
and luminosities.

\medskip
\epsfxsize=8truecm
\epsfbox{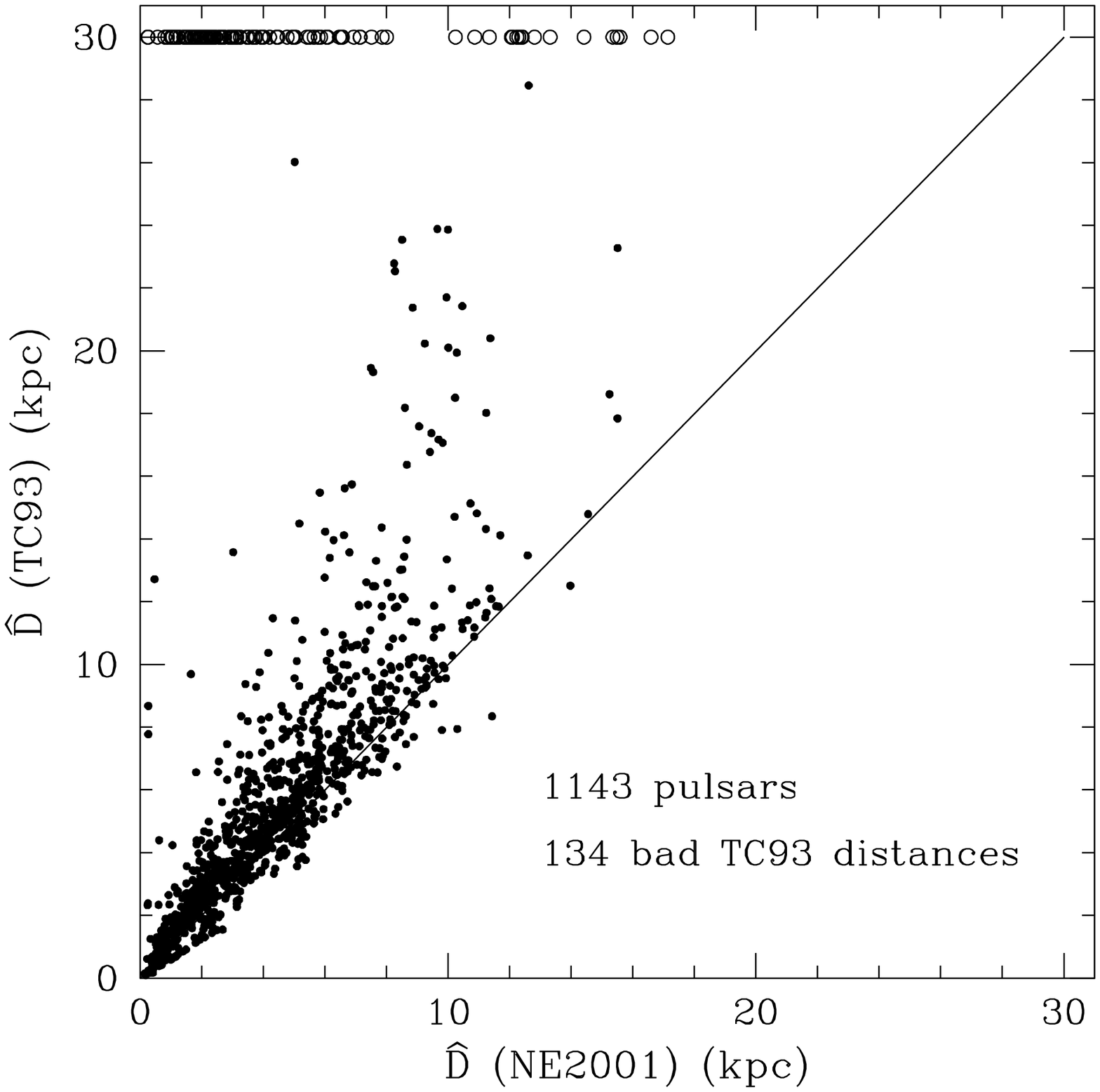}
\figcaption{
\label{fig:d2001vsdtc93}
Distance estimates from NE2001 plotted against those from
TC93.   Filled circles designate objects for which
both models yield a distance estimate.  Open circles
plotted at the top of the frame designate the 134 pulsars
for which TC93 gives only  lower distance bounds.
The solid line shows equality of the two distance estimates.
}

\subsection{Applications}\label{sec:applications}

To illustrate model predictions we show the model DM for lines of sight
in the Galactic plane ($b=0$) in Figure~\ref{fig:gplane.dm}.   
Figure~\ref{fig:contour.dm} shows DM plotted against Galactic coordinates
in an Aitoff project when integrating the electron density to infinite 
distance.   A similar pair of plots for \SM\ is shown in
Figures~\ref{fig:gplane.sm} and \ref{fig:contour.sm}. 

\medskip
\epsfxsize=8truecm
\epsfbox{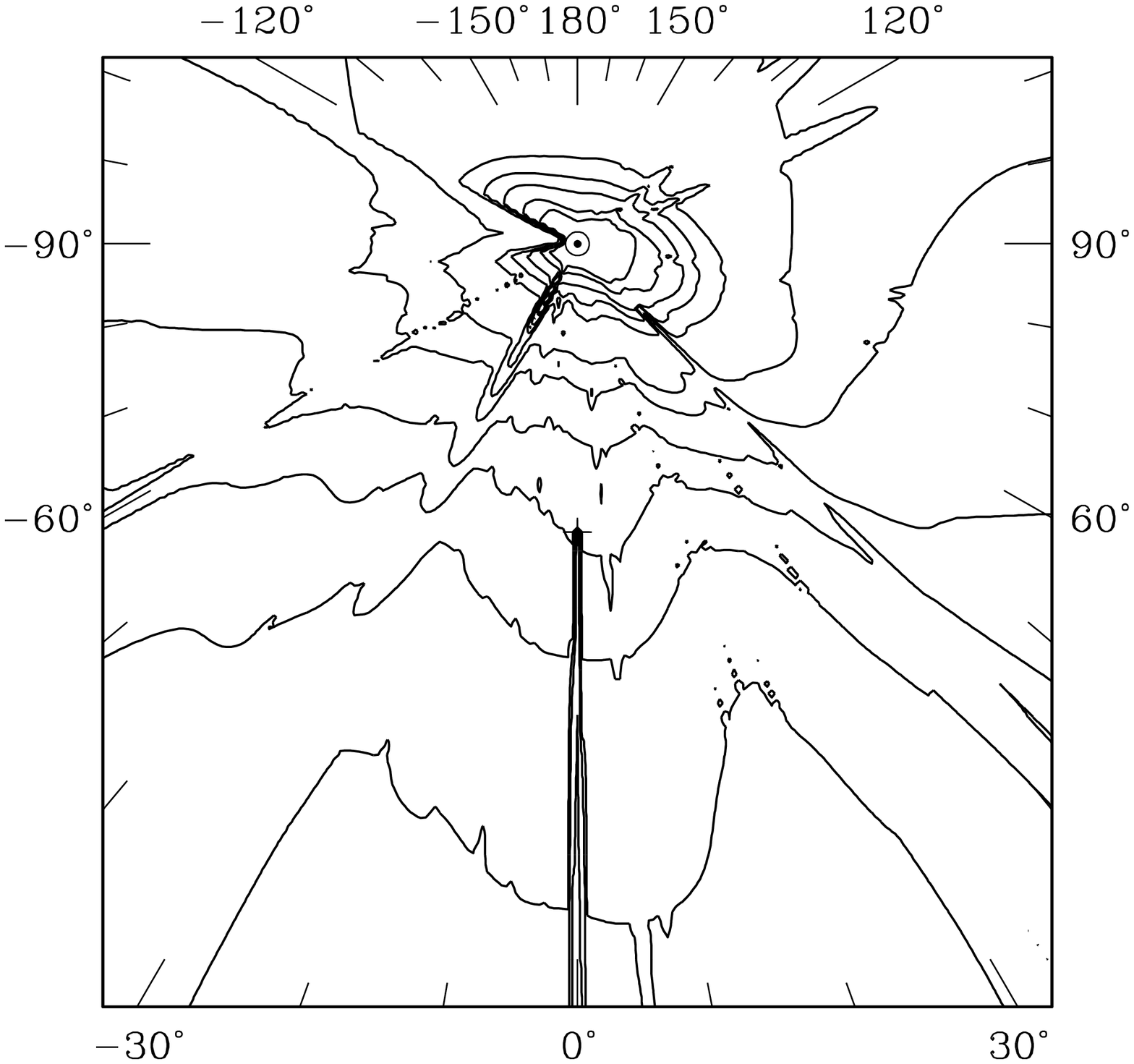}
\figcaption{
\label{fig:gplane.dm}
Contours of \DM\ plotted on the Galactic plane.
Contours are at 20, 30, 50, 70, 100, 200, 300, 500, 700, 1000, 
1500, 2000, 3000 and 4000 pc cm$^{-3}$, 
with the lowest contour nearest the Sun ($\odot$), assumed to be 
8.5 kpc from the Galactic center
(+ symbol at plot center).
}

\medskip
\epsfxsize=8truecm
\epsfbox[35 259 577 559]{figI.15.ps}
\figcaption{
\label{fig:contour.dm}
Contours of \DM\ integrated to infinite distance and
plotted against Galactic latitude and longitude on an
Aitoff projection with the Galactic center in the middle
and negative longitudes to the right. 
Contours are at $4000/2^{n}$ pc cm$^{-3}$ for $n=0,1, \ldots, 7$,
with the lowest contour at the Galactic poles.
}

\medskip
\epsfxsize=8truecm
\epsfbox{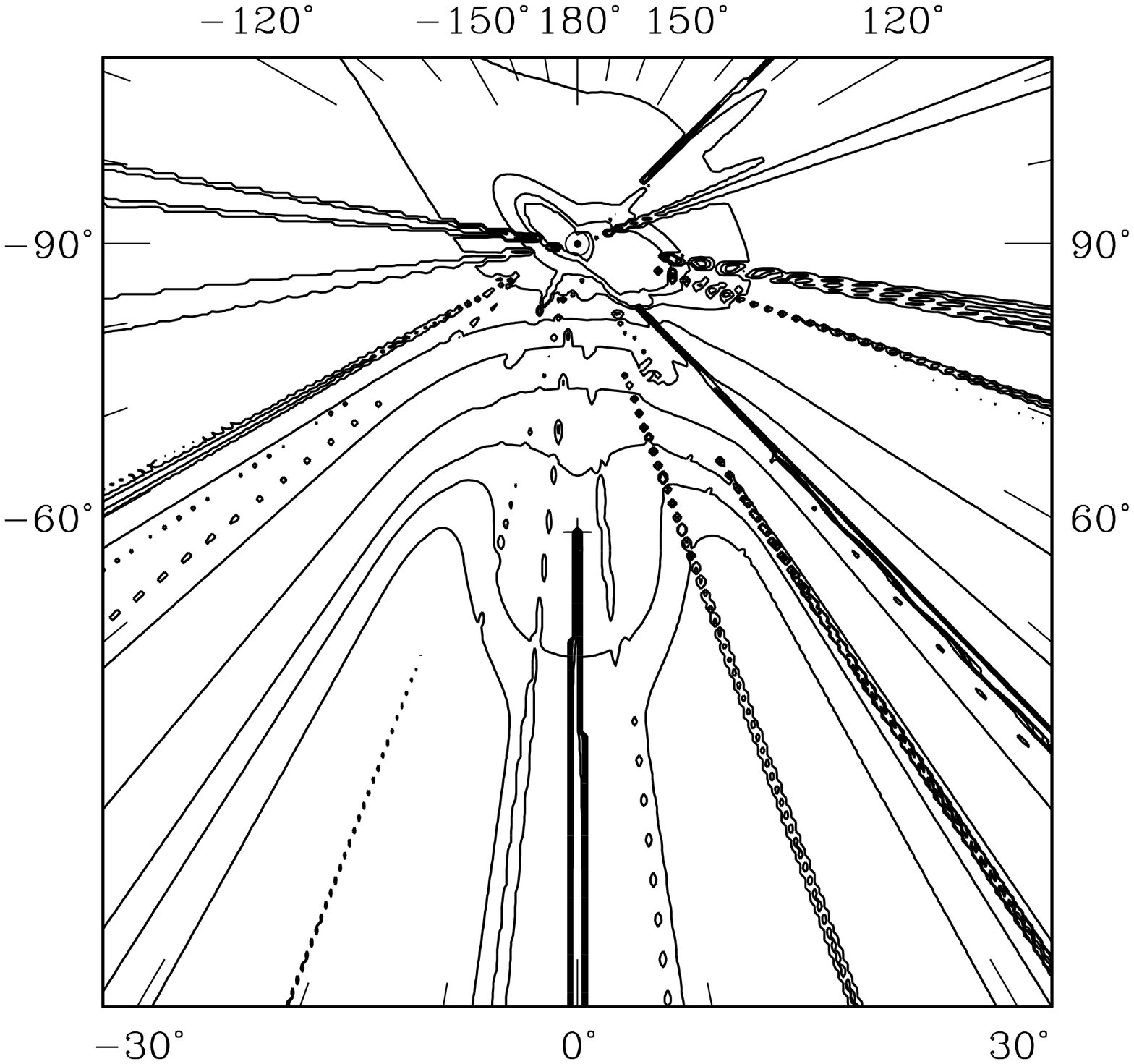}
\figcaption{
\label{fig:gplane.sm}
Contours of $\log \SM$ plotted on the Galactic plane.
Contours are at $\log\SM = 
 -5, -4, -3, -3,  -2, -1,  0, 0.5 0.67, 0.83, 1, 2, 3, 4, 5,$ and 6 
kpc m$^{-20/3}$,
chosen to bring out salient features.   The lowest contour is near the
Sun ($\odot$).   \SM\ is influenced much more than \DM\ by small
scale features in the model.
}

\medskip
\epsfxsize=8truecm
\epsfbox[35 259 577 559]{figI.17.ps}
\figcaption{
\label{fig:contour.sm}
Contours of $\log \SM$ integrated to infinite distance and
plotted against Galactic latitude and longitude on an
Aitoff projection with the Galactic center in the middle
and negative longitudes to the right. 
Contours are at $4-n/2$ kpc m$^{-20/3}$ for $n=0,1, \ldots, 16$,
with the lowest contour at the North Galactic pole.
}

\section{Summary \& Conclusions}\label{sec:discussion}

We have presented a new model, NE2001, for the Galactic distribution
of free electrons and the fluctuations within it.  As observational
constraints we make use of pulsar dispersion measures and distances
and radio-wave scattering measurements available at the end of 2001
(hence its name), and we are guided by multi-wavelength observations
of the Galaxy, particularly of the local interstellar medium.

Building on the Taylor-Cordes model (TC93), we model the free electron
distribution as composed of three large-scale components, a thick
disk, thin disk, and spiral arms.  Since the publication of the TC93
model, though, the number of available data have expanded greatly
(e.g., the number of pulsar DMs available is now approximately double
the number available to TC93).  With this larger data set, it is clear
that ``smooth,'' large-scale components are insufficient to produce a
realistic description of the electron density distribution.  We must
take into account the distribution of electrons in the local ISM, and
we require ``clumps'' and ``voids'' of electrons, mesoscale structures
distributed throughout the Galaxy on a large number of lines of sight
in order to produce reasonable agreement with the observations.

Tables~\ref{tab:ne2001}--\ref{tab:lism} summarize the model and the
best-fitting parameters of the large-scale and local ISM components.
Tables~\ref{tab:clumpsX}--\ref{tab:voids} lists the lines of sight
requiring clumps or voids and the relevant parameter for each clump or 
void.  Figure~\ref{fig:grayscale} shows the model electron density in
the Galactic plane.

We used an iterative likelihood method to find the best-fitting model
parameters.  Our focus here has been on the exposition of the model.
In a companion paper (Cordes \& Lazio 2002b) we describe details
of the fitting procedure and  discuss those lines of
sight that require a clump or
void or are otherwise problematic. 
We also discuss possible alternative  fitting approaches and models.

Our model, NE2001, improves upon the TC93 model.  First, the distance
estimates obtained from the model agree with available distance
constraints for nearly all pulsars with such constraints.  Second,
none of the parameters of the large-scale components are indeterminate
(e.g., as was the case with the thick disk for TC93).  The cost of
these improvements has been an increase in the complexity of the
model, particularly with respect to the number and location of clumps
and voids.  We believe, however, that this additional complexity is
justified both by the quantity of data and because it is astrophysically
reasonable.  A small
number of pulsars or extragalactic sources has been known for some
time to have anomalously large DMs or scattering properties or both
due to intervening \ion{H}{2} regions or supernova remnants.

While we consider NE2001 to be an improvement over TC93, we foresee a
number of probable developments that will allow future generations
of electron density models.
We group these improvements into increases in the quantity
of data and improvements in the modeling technique.  Perhaps most important
will be an increase in the number of
pulsar parallaxes, both from pulse timing methods applied 
to millisecond pulsars 
and from large programs  using very long baseline interferometry (VLBI).
Also, using pulsar luminosities 
to estimate distances should become feasible once beaming of pulsar
radiation becomes better understood and despite the fact that the 
luminosity function for radio emission from pulsars is very broad.
DM-independent
distances provide crucial calibration information for NE2001 or any
successors, and we regard it as likely that the number of pulsars with
DM-independent distances will double in the next few years.  We have
made use of only the positions and DMs of pulsars discovered in the
Parkes multibeam sample.  Efforts are underway to measure the
scattering along the lines of sight to many of these pulsars, which
could increase the number of lines of sight with measured SMs by
roughly 50\% or more.  The advent of the Green Bank Telescope and the
refurbished Arecibo telescope suggest the possibility of conducting a
northern hemisphere equivalent of the Parkes multibeam sample, which
could increase the number of pulsars by at least another 50\%.

Possible improvements in modeling include the adoption of a  more realistic
location for the Sun.  NE2001 places the Sun in the Galactic plane at
a Galactocentric distance of~8.5~kpc, though a smaller distance from
the Galactic center ($\approx 7.1$~kpc) and a modest offset from the
plane ($\approx 20$~pc) seem warranted.  Although we have provided a
formalism for comparing DM and SM to EM, we have made little use of
it.  Future work to include observational constraints on EM, e.g.,
from H$\alpha$ surveys, has the potential of producing a better model
for the local ISM.

Finally, Figure~\ref{fig:dmvsl} suggests that the large-scale
structure of the Galaxy could be determined \textit{ab
inito}, provided that a sufficient number of lines of sight exist.
Rather than imposing a large-scale structure as done both here and
previously, the presence and location of large-scale components,
particularly the spiral arms, could be determined directly
from the population of pulsars.  Future pulsar
surveys may yield the number of lines of sight required to employ
this approach.

\acknowledgments

We thank 
Z. Arzoumanian, R. Bhat, F.~Camilo, S. Chatterjee,
D. Chernoff, M. Goss, Y. Gupta, S. Johnston, 
F. J. Lockman, R. N. Manchester, and B. Rickett 
for useful discussions.  Our research is
supported by NSF grant AST~9819931 and by the National Astronomy and
Ionosphere Center, which operates the Arecibo Observatory under
cooperative agreement with the \hbox{NSF}.  Basic research in radio
astronomy at the NRL is supported by the Office of Naval Research.


\begin{deluxetable}{lrrrrrr}
\tabletypesize{\footnotesize}
\tablecolumns{7}
\tablewidth{0pc}
\tablecaption{Data Summary \label{tab:summary}}
\tablehead{
\\
	& \multicolumn{6}{c}{Number of Measurements} \\
     	& \multicolumn{6}{c}{} \\
\cline{2-7} \\
  \colhead{Work} & \colhead{D$^{\tablenotemark{a}}$} & \colhead{DM} & \colhead{$\taud$} & \colhead{$\dnud$} & \colhead{$\theta_{d,\rm Gal}$}  & \colhead{$\theta_{d,\rm xgal}$} \\
}
\startdata
TC93  		& 74 	& 553 	& 120  	& 74  	& 38 	& 42 \\ 
This paper	& 112 	& 1143 & 159 	& 110 	& 118	& 94  \\
\enddata
\tablenotetext{a}{The number of distances is the number of distinct
lines of sight.  A globular cluster with multiple pulsars is counted 
as only one line of sight.}
\end{deluxetable}

\begin{deluxetable}{lllcl}
\tabletypesize{\footnotesize}
\tablewidth{7.0in}
\tablecaption{\label{tab:ne2001}
	NE2001 Model Components \break
	  $n_e(\xvec) = (1-\wvoids) 
		\{ 
		   (1 - \wlism) \left [\negal(\xvec) + \negc(\xvec)\right ]
 	  		+ \wlism   \nelism(\xvec) 
		\}
		+ \wvoids \nevoids(\xvec)
		+ \neclumps(\xvec)
	  $
}
\tablehead{
\\
  \colhead{Component} &
  \colhead{Functional Form} &
  \colhead{Parameters} &
  \colhead{No. Parameters} &
  \colhead{Comments} 
\\
}
\startdata
\\
Smooth Components &
       \multicolumn{4}{l}
	{
	  $\negal(\xvec) =
          \left [ n_1\, G_1(r,z)  +  n_2\, G_2(r,z)  + n_a G_a(\xvec) \right ]$
	}
	\\	
	\\
\multicolumn{1}{r}{Thick Disk} &
	$n_1 G_1(r,z) = n_1 g_1(r) h(z/H_1)$ &
	$n_1, H_1, A_1, F_1$ &
	4 &
	\\
	\\
\multicolumn{1}{r}{Thin Disk} &
	$n_2 G_2(r,z) = n_2 g_2(r) h(z/H_2)$ &
	$n_2, H_2, A_2, F_2$ &
	4 &
	\\
	\\
\multicolumn{1}{r}{Spiral Arms} &
	$n_a G_a(\xvec)$ &
	$f_j n_a, h_j H_a, w_j w_a,  F_j$ 
	& 20
	\\
	&& $j=1,\ldots,5$
	\\
	\\
Galactic Center ($\negc$) &
	${\negc}_0 e^{-[\delta\rperp^2/R_{\rm GC}^2 + 
		        (z-z_{\rm GC})^2 / H_{\rm GC}^2]}$ &
	${\negc}_0, R_{\rm GC}, h_{\rm GC}$ & 
	3 &
	\\
	& $\delta\rperp^2 = (x-x_{\rm GC})^2 + (y - y_{\rm GC})^2$ \\
	\\
Local ISM ($\nelism$) &
	$\nelism(\xvec)$, $\Flism(\xvec)$, $\wlism(\xvec)$ &   
	See Table \ref{tab:lism} & 
	36 & Excludes Gum, Vela 
	\\
	& See below \& Appendix \ref{app:LISM} &  &	
	\\
	\\
Clumps ($\neclumps$)&
	$\displaystyle
	\sum_{j=1}^{\nclumps}
	        {n_c}_j 
		e^{ -\vert \xvec - {\xvec_c}_j\vert^2 /{r_c}_j^2} 
		{t_c}_j(\xvec) $ & 
        $\nclumps$ & $6\nclumps+1$ & Includes Gum, Vela 	
	\\
	&& ${n_c}_j, {\xvec_c}_j, {r_c}_j, {F_c}_j$ & (6/clump) & 
	\\
	\\
Voids ($\nevoids$)  &
	$\displaystyle
	\sum_{j=1}^{\nvoids}
	        {n_v}_j 
		g_v(\xvec; {\thetavec_v}_j)
		{t_v}_j(\xvec) $ & 
        $\nvoids$ & $8\nvoids+1$ & 
	\\
	&&
	${n_v}_j, {\xvec_v}_j, {\thetavec_v}_j, {F_v}_j$ & 
	(8/void)   
	&
	\\
	\\
\hline
	\\
Functions: \\
      \multicolumn{4}{l}
	{
	   \quad\quad
	   $h(x) = {\rm sech}^2(x)$ 
	}
	\\
      \multicolumn{4}{l}
	{
	   \quad\quad
	   $U(x)$ = {\rm unit step function} 
	}
	\\
      \multicolumn{4}{l}
	{
	    \quad\quad
	    $g_1(r) = [\cos(\pi r/2A_1)/\cos(\pi\rsun/2A_1)] U(r-A_1)$ 
	}
	\\
      \multicolumn{4}{l}
	{
	    \quad\quad
	    $g_2(r) = \exp(-(r-A_a)^2/A_a^2) U(r)$ 
	}
	\\
      \multicolumn{3}{l}
	{
	    \quad\quad
	    $G_a(\xvec) = \sum_j f_j {g_a}_j(r, s_j(\xvec)/w_jw_a)h(z/h_j h_a)$ 
	}
	&
	\multicolumn{2}{r}{$s_j(\xvec)$ tabulated}
	\\
      \multicolumn{4}{l}
	{
	   \quad\quad
	   ${g_a}_j(\xvec) = 
              e^{-(s_j(\xvec)/w_jw_a)^2} {\rm sech}^2(r-A_a)/2 U(r-A_a)$
	}
	\\
      \multicolumn{4}{l}
	{
	   \quad\quad
	
	$ \nelism(\xvec) = 
	 (1-\wlhb)
		\{
		   (1-\wloopI) [(1-\wlsb)\neldr(\xvec) + \wlsb\nelsb(\xvec)] + 
	 		\wloopI \neloopI(\xvec) \} + \wlhb\nelhb(\xvec)]$
	}
	\\
      \multicolumn{4}{l}
	{
	   \quad\quad
	
	$ \Flism(\xvec) = 
	 (1-\wlhb)
		\{
		   (1-\wloopI) [(1-\wlsb)\Fldr(\xvec) + \wlsb\Flsb(\xvec)] + 
	             \wloopI \FloopI(\xvec) \} + \wlhb\Flhb(\xvec)]$
	}
	\\
      \multicolumn{3}{l}
	{
	  \quad\quad
		$\wlism(\xvec) = \mathrm{max} [
			\wldr(\xvec),\wlhb(\xvec),\wlsb(\xvec),\wloopI(\xvec)]
			= (0,1)
		$
	}
	&
	\multicolumn{2}{r} {LISM weight function}
	\\
      \multicolumn{3}{l}
	{
	  \quad\quad
   	  ${t_c}_j(\xvec) = 
		[1 - {e_c}_j U(\vert \xvec - {\xvec_c}_j\vert - {r_c}_j)],$ 
	    	\quad ${e_c}_j = (0,1)$
	}
	&
	\multicolumn{2}{r} {truncation function}
	\\
      \multicolumn{4}{l}
	{
	  \quad\quad
	  $g_v(\xvec,{\thetavec_v}_j)$ = elliptical gaussian = $\exp(-Q(\xvec-{\xvec_c}_j)$, \quad 
		 ${\thetavec_v}_j = (a_j, b_j, c_j, \theta_{y_j}, \theta_{z_j})$
	}
	&
	\\
      \multicolumn{3}{l}
	{
	  \quad\quad
   	  ${t_v}_j(\xvec) = 
		[1 - {e_v}_j U(Q-1)],$ 
	    	\quad ${e_v}_j = (0,1)$
	}
	&
	\multicolumn{2}{r} {truncation function}
	\\
      \multicolumn{3}{l}
	{
	  \quad\quad
		$Q = 
		(\xvec - {\xvec_c}_j)^{\dagger}
		{\bf V^{-1}}
		(\xvec - {\xvec_c}_j)$, 
		$V$ = rotation matrix 
        }
        &
	\multicolumn{2}{r} {ellipsoidal quadratic form}
	\\
	\multicolumn{3}{l} {
	  \quad\quad
	  $\wvoids = (0,1)$
	}
	&
	\multicolumn{2}{r} {weight function for voids}
	\\
Weight functions: \\
      \multicolumn{4}{l}
	{
	   \quad\quad
	   $\wvoids, \wlism, \wldr, \wlhb, \wlsb, \wloopI$ switch components
	   on and off.
	}
\\
Truncation functions: \\
      \multicolumn{4}{l}
	{
	   \quad\quad
	   ${t_c}_j$, ${t_v}_j$ truncate component at 1/e point if
	   ${e_c}_j$, ${e_v}_j$ = 1. 
	}
\\
\\
\enddata
\end{deluxetable}{}

\begin{deluxetable}{lll}
\tabletypesize{\footnotesize}
\tablewidth{0pc}
\tablecolumns{3}
\tablecaption{Parameters of Large Scale Components of 
	TC93 and NE2001 \label{tab:large}}
\tablehead{
\\
\colhead{Parameter} & \colhead{TC93 Values} & \colhead{NE2001 Values$^a$} \\
}
\startdata
$n_1 h_1$ (cm$^{-3}\,$kpc)\dots & $0.0165\pm0.0006$ & 0.033 \\
$h_1$ (kpc) \dotfill & $0.88\pm0.06$ &  0.95 \\
$A_1$ (kpc) \dotfill & $\ga20$ & 17 \\
$F_1$ \dotfill & $0.36^{+0.30}_{-0.10}$ & 0.20  \\
$n_2$ (cm$^{-3}$) \dotfill & $0.10\pm0.03$ & 0.090  \\
$h_2$ (kpc) \dotfill & $0.15\pm0.05$ & 0.14 \\
$A_2$ (kpc) \dotfill & $3.7\pm0.3$ &  3.7\\
$F_2$ \dotfill & $43^{+30}_{-13}$ & 110 \\
$n_a f_j$ (cm$^{-3}$) \dotfill & $0.084\pm0.008$ & 0.030 $\times ( 0.50, 1.2, 1.3, 1.0, 0.25)$ \\
$h_a h_j$ (kpc) \dotfill & $0.3\pm0.1$ & 0.25  $\times ( 1.0, 0.8, 1.3, 1.5, 1.0 )$ \\
$w_a w_j$ (kpc) \dotfill & $0.3$ &  0.6 $\times ( 1, 1.5, 1, 0.8, 1)$\\
$A_a$ (kpc) \dotfill & $8.5$ &  11.0\\
$F_a F_j$ \dotfill & $6^{+5}_{-2}$ & 10   $\times ( 1.1, 0.3, 0.4, 1.5, 0.3)$ \\
$n_G$ (cm$^{-3}$)\dotfill & 0.25 & \nodata \\
$F_G$ \dotfill &     0.0  	& \nodata \\
$n_{GC}$ (cm$^{-3}$)\dotfill    &\nodata &  10.0\\
$F_{GC}$ \dotfill & \nodata  & $5\times 10^4$\\
\\
\enddata
\end{deluxetable}

\begin{deluxetable}{lrrrrrrrrrrr}
\tabletypesize{\footnotesize}
\tablewidth{0pc}
\tablecolumns{12}
\tablecaption{Local ISM Components \label{tab:lism}}
\tablehead{
\\
\colhead{Component}
     & \multicolumn{3}{c}{Location} && \multicolumn{4}{c}{Size \& Shape} 
     && \multicolumn{2}{c} {Density}     \\
\\
\cline{2-4} \cline{6-9}  \cline{11-12} \\ 
     & \colhead{$\xbar$\quad} 
     & \colhead{$\ybar$\quad} 
     & \colhead{$\zbar$\quad}
&& 
	\colhead{a} & \colhead{b} & \colhead{c} & 
        \colhead{$\theta^{\tablenotemark{a}}$} && \colhead{$n_e$} & 
        \colhead{$F$} \\ 
     & \colhead{(kpc)} 
     & \colhead{(kpc)} 
     & \colhead{(kpc)} &
     & \colhead{(kpc)} 
     & \colhead{(kpc)} 
     & \colhead{(kpc)} 
     & \colhead{(deg)} &
     & \colhead{(cm$^{-3}$)} \\
}
\startdata
\\
LDR & 1.36  & 8.06 & 0.0  && 1.5 & 0.75 & 0.5  & $-24.8$ && 0.012 & 0.1 \\
LHB & 0.01    & 8.45 & 0.17 && 0.085 & 0.1 & 0.33 & 15 && 0.005 & 0.01 \\  
LSB & $-0.75$ & 9.0 & $-0.05$ && 1.05 & 0.425 & 0.325 & 139 && 0.016 & 0.01 \\  
\\
\tableline
\\
\colhead{Loop I (NPS)}
     & \multicolumn{3}{c}{Location} && \multicolumn{2}{c}{Size} 
     && \multicolumn{4}{c} {Density}     \\
\\
\cline{2-4} \cline{6-7}  \cline{9-12} \\ 
     & \colhead{$\xbar$} 
     & \colhead{$\ybar$}
     & \colhead{$\zbar$} &
     & \colhead{r} 
     & \colhead{$\Delta r$} &
     & \colhead{$n_e$} 
     & \colhead{$\Delta n_e$} 
     & \colhead{$F_{\rm vol}$} 
     & \colhead{$F_{\rm shell}$} \\ 
     & \colhead{(kpc)} 
     & \colhead{(kpc)}  
     & \colhead{(kpc)} 
     && \colhead{(kpc)} 
     & \colhead{(kpc)}   &
     & \colhead{(cm$^{-3}$)} 
     & \colhead{(cm$^{-3}$)} \\
\\
\tableline
\\
        & $-0.045$ & 8.4 & 0.07 && 0.14 & 0.03  && 0.0125 & 0.010 & 0.20 &  0.01 \\
\\
\enddata
\tablenotetext{a}
{For LSB and LDR, $\theta$ is the angle of the major axis of
the ellipsoid with respect to the the $x$ axis
($\ell = 90^{\circ}$), increasing counterclockwise looking down
on the Galactic plane. 
For the LHB, $\theta$ is measured from the $z$ axis and describes the slant of
the axis of the cylinder in the y-z plane.   Loop I has a hemispherical shape
with internal volume and shell component 
that contributes only for $z\ge 0$.}
\end{deluxetable}

\begin{deluxetable}{lrrrrrr}
\tabletypesize{\footnotesize}
\tablewidth{0pc}
\tablecolumns{5}
\footnotesize
\tablecaption{Clump Parameters for Lines of Sight to Extragalactic Sources
     \label{tab:clumpsX}}
\tablehead{
\\
\multicolumn{1}{c} {LOS} & 
\multicolumn{1}{c} {$\ell$} & 
\multicolumn{1}{c} {$b$} & 
\multicolumn{1}{c} {$\dc$} & 
\multicolumn{1}{c} {$\log\SM_c$} & 
\multicolumn{1}{c} {$\DM_c$} &  
\multicolumn{1}{c} {$r_c$}  \\
&
\multicolumn{1}{c} {(deg)} & 
\multicolumn{1}{c} {(deg)} & 
\multicolumn{1}{c} {(kpc)} & 
\multicolumn{1}{c} {(kpc m$^{-20/3}$)} &  
\multicolumn{1}{c} {(pc cm$^{-3}$)} &  
\multicolumn{1}{c} {(kpc)} \\  
\\
}
\startdata
\\
${\rm B1741-312}$   &  $  -2.14$  &  $  -1.00$  &  $  8.50$  &   $ 1.40$ \quad\quad\quad   &  $    89$ \quad\quad\quad  &  $0.01$ \quad   \\ 
${\rm 1905+079}$   &  $  41.91$  &  $   0.09$  &  $  8.00$  &   $ 1.55$ \quad\quad\quad   &  $   106$ \quad\quad\quad  &  $0.01$ \quad   \\ 
${\rm 2008+33D}$   &  $  71.16$  &  $  -0.09$  &  $  2.35$  &   $ 0.35$ \quad\quad\quad   &  $    27$ \quad\quad\quad  &  $0.01$ \quad   \\ 
${\rm 2021+317}$   &  $  71.40$  &  $  -3.10$  &  $  2.35$  &   $-0.91$ \quad\quad\quad   &  $     6$ \quad\quad\quad  &  $0.01$ \quad   \\ 
${\rm 2023+336}$   &  $  73.10$  &  $  -2.40$  &  $  2.35$  &   $-0.20$ \quad\quad\quad   &  $    14$ \quad\quad\quad  &  $0.01$ \quad   \\ 
${\rm 2014+358}$   &  $  74.04$  &  $   0.36$  &  $  2.35$  &   $-0.09$ \quad\quad\quad   &  $    16$ \quad\quad\quad  &  $0.01$ \quad   \\ 
${\rm 2020+351}$   &  $  74.14$  &  $  -1.01$  &  $  2.35$  &   $ 0.68$ \quad\quad\quad   &  $    39$ \quad\quad\quad  &  $0.01$ \quad   \\ 
${\rm 2005+372}$   &  $  74.18$  &  $   2.61$  &  $  2.35$  &   $ 0.56$ \quad\quad\quad   &  $    34$ \quad\quad\quad  &  $0.01$ \quad   \\ 
${\rm 2048+313}$   &  $  74.60$  &  $  -8.04$  &  $  2.35$  &   $-0.09$ \quad\quad\quad   &  $    16$ \quad\quad\quad  &  $0.01$ \quad   \\ 
${\rm 2013+370}$   &  $  74.90$  &  $   1.20$  &  $  2.20$  &   $-0.45$ \quad\quad\quad   &  $    11$ \quad\quad\quad  &  $0.01$ \quad   \\ 
${\rm 2012+383}$   &  $  75.78$  &  $   2.19$  &  $  2.35$  &   $-0.05$ \quad\quad\quad   &  $    17$ \quad\quad\quad  &  $0.01$ \quad   \\ 
${\rm 2005+403}$   &  $  76.80$  &  $   4.30$  &  $  2.35$  &   $-0.00$ \quad\quad\quad   &  $    18$ \quad\quad\quad  &  $0.01$ \quad   \\ 
${\rm 2050+364}$   &  $  78.90$  &  $  -5.10$  &  $  2.35$  &   $-1.40$ \quad\quad\quad   &  $     4$ \quad\quad\quad  &  $0.02$ \quad   \\ 
${\rm 0241+622}$   &  $ 135.70$  &  $   2.40$  &  $  2.20$  &   $-0.09$ \quad\quad\quad   &  $    16$ \quad\quad\quad  &  $0.01$ \quad   \\ 

\\
\enddata
\end{deluxetable}

\begin{deluxetable}{lrrrrrr}
\tabletypesize{\footnotesize}
\tablewidth{0pc}  
\tablecolumns{5}
\tablecaption{Clump Parameters for Galactic, Non-Pulsar Lines of Sight
	\label{tab:clumpsG}}
\tablehead{
\\
\multicolumn{1}{c}{LOS} & 
\multicolumn{1}{c}{$\ell$} & 
\multicolumn{1}{c}{$b$} & 
\multicolumn{1}{c}{$\dc$} & 
\multicolumn{1}{c}{$\log\SM_c$} & 
\multicolumn{1}{c}{$\DM_c$} & 
\multicolumn{1}{c}{$r_c$} \\ 
&
\multicolumn{1}{c}{(deg)} & 
\multicolumn{1}{c}{(deg)} & 
\multicolumn{1}{c}{(kpc)} & 
\multicolumn{1}{c}{(kpc m$^{-20/3})$} & 
\multicolumn{1}{c}{(pc cm$^{-3})$} & 
\multicolumn{1}{c}{(kpc)} \\ 
\\
}
\startdata
\\
${\rm MonR2}$   &  $ -146.30$  &  $ -12.60$  &  $  0.41$  &   $-0.96$ \quad\quad\quad    &  $     6$ \quad\quad\quad  &  $0.01$  \quad   \\ 
${\rm NGC6334N}^a$   &  $   -8.80$  &  $   0.65$  &  $  1.67$  &   $ 3.61$ \quad\quad\quad    &  $   142$ \quad\quad\quad  &  $0.01$  \quad   \\ 
${\rm OH353.298}$   &  $   -6.70$  &  $  -1.54$  &  $  8.40$  &   $ 6.35$ \quad\quad\quad    &  $   886$ \quad\quad\quad  &  $0.05$  \quad   \\ 
${\rm OH355}$   &  $   -4.23$  &  $  -1.75$  &  $  8.40$  &   $ 6.46$ \quad\quad\quad    &  $   886$ \quad\quad\quad  &  $0.05$  \quad   \\ 
${\rm OH357.849}$   &  $   -2.15$  &  $   9.74$  &  $  8.40$  &   $ 5.75$ \quad\quad\quad    &  $   886$ \quad\quad\quad  &  $0.05$  \quad   \\ 
${\rm OH359.140}$   &  $   -0.86$  &  $   1.14$  &  $  8.40$  &   $ 7.00$ \quad\quad\quad    &  $   886$ \quad\quad\quad  &  $0.05$  \quad   \\ 
${\rm OH359.540}$   &  $   -0.44$  &  $   1.29$  &  $  8.40$  &   $ 4.79$ \quad\quad\quad    &  $   886$ \quad\quad\quad  &  $0.05$  \quad   \\ 
${\rm OH000.125+5.111}$   &  $    0.12$  &  $   5.11$  &  $  8.40$  &   $ 4.60$ \quad\quad\quad    &  $   886$ \quad\quad\quad  &  $0.05$  \quad   \\ 
${\rm OH000.892+1.342}$   &  $    0.89$  &  $   1.34$  &  $  8.40$  &   $ 4.95$ \quad\quad\quad    &  $   886$ \quad\quad\quad  &  $0.05$  \quad   \\ 
${\rm OH20.1-0.1}$   &  $   20.07$  &  $  -0.09$  &  $  4.00$  &   $ 3.56$ \quad\quad\quad    &  $   177$ \quad\quad\quad  &  $0.01$  \quad   \\ 
${\rm OH35.2-1.7}$   &  $   35.20$  &  $  -1.73$  &  $  2.80$  &   $ 0.30$ \quad\quad\quad    &  $    35$ \quad\quad\quad  &  $0.01$  \quad   \\ 
${\rm OH40.6-0.2}$   &  $   40.62$  &  $  -0.14$  &  $  2.20$  &   $ 3.75$ \quad\quad\quad    &  $    53$ \quad\quad\quad  &  $0.00$  \quad   \\ 
${\rm W49N}$   &  $   43.16$  &  $   0.01$  &  $  8.50$  &   $ 1.69$ \quad\quad\quad    &  $   124$ \quad\quad\quad  &  $0.02$  \quad   \\ 
${\rm OH43.80-0.13}$   &  $   43.78$  &  $  -0.14$  &  $  2.60$  &   $ 2.16$ \quad\quad\quad    &  $   213$ \quad\quad\quad  &  $0.02$  \quad   \\ 
${\rm CygX-3}$   &  $   79.85$  &  $   0.70$  &  $  2.35$  &   $ 1.69$ \quad\quad\quad    &  $   124$ \quad\quad\quad  &  $0.02$  \quad   \\ 
${\rm W75S}$   &  $   81.80$  &  $   0.64$  &  $  2.40$  &   $ 1.69$ \quad\quad\quad    &  $   124$ \quad\quad\quad  &  $0.02$  \quad   \\ 
${\rm CepA}$   &  $  109.87$  &  $   2.10$  &  $  0.60$  &   $ 0.16$ \quad\quad\quad    &  $    21$ \quad\quad\quad  &  $0.02$  \quad   \\ 
${\rm NGC7538}$   &  $  111.54$  &  $   0.77$  &  $  3.40$  &   $-0.00$ \quad\quad\quad    &  $    18$ \quad\quad\quad  &  $0.02$  \quad   \\ 
${\rm W3OH}$   &  $  133.95$  &  $   1.06$  &  $  2.20$  &   $ 3.80$ \quad\quad\quad    &  $   354$ \quad\quad\quad  &  $0.02$  \quad   \\ 

\\
\enddata
\tablenotetext{a}{The clump toward NGC6334N also affects the 
strongly scattered extragalactic source, NGC6334B (see Paper II).
}
\end{deluxetable}

\begin{deluxetable}{lrrrrrr}
\tabletypesize{\footnotesize}
\tablecolumns{7}
\tablewidth{0pc}
\tablecaption{
	Clumps for Pulsar Lines of Sight with 
	$\DMc > 20$ pc cm$^{-3}$ or $\log\SMc > 0$ 
	\label{tab:clumpsP.short}
}
\tablehead{
\\
\multicolumn{1}{l}{ LOS} & 
\multicolumn{1}{c}{$\ell$} & 
\multicolumn{1}{c}{$b$} &  
\multicolumn{1}{c}{$\dc$} & 
\multicolumn{1}{c}{$\DM_c$} &  
\multicolumn{1}{c}{$\log\SM_c$} & 
\multicolumn{1}{c}{$r_c$}  \\
&
\multicolumn{1}{c}{ (deg)} & 
\multicolumn{1}{c}{(deg)} & 
\multicolumn{1}{c}{(kpc)} & 
\multicolumn{1}{c}{(pc cm$^{-3}$)} & 
\multicolumn{1}{c}{(kpc m$^{-20/3})$} & 
\multicolumn{1}{c}{(kpc)}  \\ 
\\
}
\startdata
${\rm 0611+22}$  &  $ -171.21$  &  $   2.40$  &  $   0.65$  &  $   23.2$ \quad\quad  &  $\cdots$ \quad\quad\quad  &  $ 0.01$  \\ 
${\rm GumI}$  &  $ -100.00$  &  $  -1.00$  &  $   0.50$  &  $  110.5$ \quad\quad  &  $\cdots$ \quad\quad\quad  &  $ 0.14$  \\ 
${\rm GumII}$  &  $ -97.20$  &  $   2.70$  &  $   0.50$  &  $   28.4$ \quad\quad  &  $-0.59$ \quad\quad\quad  &  $ 0.03$ \\ 
${\rm VelaIras}$  &  $ -96.75$  &  $  -9.00$  &  $   0.25$  &  $  269.4$ \quad\quad  &  $ 1.54$ \quad\quad\quad  &  $ 0.04$ \\ 
${\rm J1019-5749}$  &  $ -76.20$  &  $  -0.68$  &  $   4.21$  &  $  709.0$ \quad\quad  &  $ 2.60$ \quad\quad\quad  &  $ 0.01$ \\ 
${\rm J1022-5813}$  &  $ -75.70$  &  $  -0.83$  &  $   4.21$  &  $  354.5$ \quad\quad  &  $ 2.60$ \quad\quad\quad  &  $ 0.01$ \\ 
${\rm J1031-6117}$  &  $ -73.12$  &  $  -2.88$  &  $   3.00$  &  $   35.4$ \quad\quad  &  $-0.70$ \quad\quad\quad  &  $ 0.02$ \\ 
${\rm J1056-6258}$  &  $ -69.71$  &  $  -2.97$  &  $   2.20$  &  $  177.2$ \quad\quad  &  $-0.13$ \quad\quad\quad  &  $ 0.02$ \\ 
${\rm 1112-60}$  &  $ -68.56$  &  $  -0.32$  &  $   6.13$  &  $    9.9$ \quad\quad  &  $ 0.19$ \quad\quad\quad  &  $ 0.01$ \\ 
${\rm J1119-6127}$  &  $ -67.85$  &  $  -0.54$  &  $   2.30$  &  $   35.4$ \quad\quad  &  $-0.70$ \quad\quad\quad  &  $ 0.02$ \\ 
${\rm J1128-6219}$  &  $ -66.51$  &  $  -0.97$  &  $   2.30$  &  $   35.4$ \quad\quad  &  $ 0.60$ \quad\quad\quad  &  $ 0.02$ \\ 
${\rm 1131-62}$  &  $ -65.79$  &  $  -1.30$  &  $   6.51$  &  $   10.8$ \quad\quad  &  $ 0.27$ \quad\quad\quad  &  $ 0.01$ \\ 
${\rm J1201-6306}$  &  $ -62.69$  &  $  -0.79$  &  $   2.00$  &  $   78.0$ \quad\quad  &  $-0.02$ \quad\quad\quad  &  $ 0.02$ \\ 
${\rm J1216-6223}$  &  $ -61.08$  &  $   0.20$  &  $   2.10$  &  $  106.3$ \quad\quad  &  $ 0.25$ \quad\quad\quad  &  $ 0.02$ \\ 
${\rm J1305-6256}$  &  $ -55.50$  &  $  -0.12$  &  $   8.00$  &  $  354.5$ \quad\quad  &  $ 2.60$ \quad\quad\quad  &  $ 0.01$ \\ 
${\rm J1324-6146}$  &  $ -53.10$  &  $   0.85$  &  $   8.00$  &  $  354.5$ \quad\quad  &  $ 2.60$ \quad\quad\quad  &  $ 0.01$ \\ 
${\rm 1334-61}$  &  $ -51.63$  &  $   0.30$  &  $   6.50$  &  $   13.5$ \quad\quad  &  $ 0.46$ \quad\quad\quad  &  $ 0.01$ \\ 
${\rm 1338-62}$  &  $ -51.27$  &  $  -0.04$  &  $   7.21$  &  $   12.6$ \quad\quad  &  $ 0.40$ \quad\quad\quad  &  $ 0.01$ \\ 
${\rm 1430-6623}$  &  $  -47.35$  &  $  -5.40$  &  $   0.90$  &  $   33.1$ \quad\quad  &  $\cdots$ \quad\quad\quad  &  $ 0.01$  \\ 
${\rm 1508-57}$  &  $  -39.23$  &  $  -0.11$  &  $   3.00$  &  $   88.6$ \quad\quad  &  $\cdots$ \quad\quad\quad  &  $ 0.01$  \\ 
${\rm 1419-3920}$  &  $  -39.07$  &  $  20.45$  &  $   0.30$  &  $   26.1$ \quad\quad  &  $\cdots$ \quad\quad\quad  &  $ 0.01$  \\ 
${\rm 1627-47}$  &  $ -23.52$  &  $   0.62$  &  $   3.43$  &  $   33.9$ \quad\quad  &  $ 1.26$ \quad\quad\quad  &  $ 0.01$ \\ 
${\rm 1630-44}$  &  $ -21.27$  &  $   1.98$  &  $   3.38$  &  $   10.8$ \quad\quad  &  $ 0.27$ \quad\quad\quad  &  $ 0.01$ \\ 
${\rm 1641-45}$  &  $ -20.81$  &  $  -0.20$  &  $   3.40$  &  $   53.2$ \quad\quad  &  $ 0.25$ \quad\quad\quad  &  $ 0.02$ \\ 
${\rm 1643-43}$  &  $ -18.89$  &  $   0.97$  &  $   3.34$  &  $   10.8$ \quad\quad  &  $ 0.27$ \quad\quad\quad  &  $ 0.01$ \\ 
${\rm 1703-40}$  &  $ -14.28$  &  $  -0.20$  &  $   1.45$  &  $   15.2$ \quad\quad  &  $ 0.57$ \quad\quad\quad  &  $ 0.01$ \\ 
${\rm 1718-36}$  &  $   -9.07$  &  $  -0.00$  &  $   1.50$  &  $   72.0$ \quad\quad  &  $\cdots$ \quad\quad\quad  &  $ 0.01$  \\ 
${\rm 1727-33}$  &  $  -5.87$  &  $   0.09$  &  $   1.48$  &  $   21.4$ \quad\quad  &  $ 0.86$ \quad\quad\quad  &  $ 0.01$ \\ 
${\rm 1736-31}$  &  $  -2.90$  &  $  -0.22$  &  $   3.31$  &  $   17.0$ \quad\quad  &  $ 0.66$ \quad\quad\quad  &  $ 0.01$ \\ 
${\rm 1746-30}$  &  $   -0.54$  &  $  -1.24$  &  $   2.00$  &  $   77.3$ \quad\quad  &  $\cdots$ \quad\quad\quad  &  $ 0.01$  \\ 
${\rm 1807-2715}$  &  $    3.84$  &  $  -3.26$  &  $   1.50$  &  $   70.2$ \quad\quad  &  $\cdots$ \quad\quad\quad  &  $ 0.01$  \\ 
${\rm 1758-23}$  &  $   6.81$  &  $  -0.08$  &  $   6.63$  &  $   17.9$ \quad\quad  &  $ 0.71$ \quad\quad\quad  &  $ 0.01$ \\ 
${\rm 1815-14}$  &  $  16.41$  &  $   0.61$  &  $   4.83$  &  $   11.7$ \quad\quad  &  $ 0.34$ \quad\quad\quad  &  $ 0.01$ \\ 
${\rm 1820-14}$  &  $   17.25$  &  $  -0.18$  &  $   3.50$  &  $  301.3$ \quad\quad  &  $\cdots$ \quad\quad\quad  &  $ 0.01$  \\ 
${\rm 1824-10}$  &  $  21.29$  &  $   0.80$  &  $   1.92$  &  $   15.2$ \quad\quad  &  $ 0.57$ \quad\quad\quad  &  $ 0.01$ \\ 
${\rm 1836-1008}$  &  $   22.26$  &  $  -1.42$  &  $   2.00$  &  $   47.1$ \quad\quad  &  $\cdots$ \quad\quad\quad  &  $ 0.01$  \\ 
${\rm 1830-08}$  &  $   23.39$  &  $   0.06$  &  $   4.40$  &  $  141.8$ \quad\quad  &  $\cdots$ \quad\quad\quad  &  $ 0.01$  \\ 
${\rm 1832-06}$  &  $  25.09$  &  $   0.55$  &  $   2.06$  &  $   36.5$ \quad\quad  &  $ 1.32$ \quad\quad\quad  &  $ 0.01$ \\ 
${\rm 1839-04}$  &  $  28.35$  &  $   0.17$  &  $   2.23$  &  $   15.8$ \quad\quad  &  $ 0.60$ \quad\quad\quad  &  $ 0.01$ \\ 
${\rm 1849+00}$  &  $  33.30$  &  $   0.10$  &  $   7.40$  &  $  354.5$ \quad\quad  &  $ 3.60$ \quad\quad\quad  &  $ 0.02$ \\ 
${\rm 1821+05}$  &  $   34.99$  &  $   8.86$  &  $   0.44$  &  $   22.5$ \quad\quad  &  $\cdots$ \quad\quad\quad  &  $ 0.01$  \\ 
${\rm 1900+01}$  &  $  35.73$  &  $  -1.96$  &  $   2.80$  &  $  141.8$ \quad\quad  &  $ 0.11$ \quad\quad\quad  &  $ 0.04$ \\ 
${\rm 1859+07}$  &  $  40.57$  &  $   1.06$  &  $   3.50$  &  $   35.4$ \quad\quad  &  $-0.70$ \quad\quad\quad  &  $ 0.02$ \\ 
${\rm 1907+10}$  &  $   44.83$  &  $   0.99$  &  $   0.60$  &  $   34.6$ \quad\quad  &  $\cdots$ \quad\quad\quad  &  $ 0.01$  \\ 
${\rm 1914+13}$  &  $   47.58$  &  $   0.45$  &  $   4.00$  &  $   88.8$ \quad\quad  &  $\cdots$ \quad\quad\quad  &  $ 0.01$  \\ 
${\rm 2044+46}$  &  $  85.43$  &  $   2.11$  &  $   2.00$  &  $  177.2$ \quad\quad  &  $ 2.00$ \quad\quad\quad  &  $ 0.01$ \\ 
${\rm 2036+53}$  &  $  90.37$  &  $   7.31$  &  $   2.00$  &  $   88.6$ \quad\quad  &  $ 1.40$ \quad\quad\quad  &  $ 0.01$ \\ 

\\
\enddata
\end{deluxetable}

\begin{deluxetable}{lrrrrrrrrr}
\tablefontsize{\footnotesize}
\tablewidth{0pc}
\tablecolumns{11}
\tablecaption{Void Parameters \label{tab:voids}}
\tablehead{
\\
\multicolumn{1}{c}{ LOS} &  
\multicolumn{1}{c}{ $\ell$} & 
\multicolumn{1}{c}{ $b$} &  
\multicolumn{1}{c}{ $d_v$} & 
\multicolumn{1}{c}{ ${n_e}_v$} & 
\multicolumn{1}{c}{ $F_v$} &
\multicolumn{1}{c}{ $a_v$} & 
\multicolumn{1}{c}{ $b_v$} & 
\multicolumn{1}{c}{ $c_v$} & 
\multicolumn{1}{c}{ ${\theta_v}_z$}  \\
\multicolumn{1}{c}{ (1)} & 
\multicolumn{1}{c}{ (2)} & 
\multicolumn{1}{c}{ (3)} & 
\multicolumn{1}{c}{ (4)} & 
\multicolumn{1}{c}{ (5)} & 
\multicolumn{1}{c}{ (6)} & 
\multicolumn{1}{c}{ (7)} & 
\multicolumn{1}{c}{ (8)} & 
\multicolumn{1}{c}{ (9)} & 
\multicolumn{1}{c}{ (10) } \\ 
\\
}
\startdata
\\
${\rm GumIedge}$  &  $   -81.50$  &  $  -0.60$  &  $  0.50$  &  $ 0.500$  &  $ 1.00$  &  $ 0.02$  &  $ 0.02$  &  $ 0.04$  &  $ 9.00$  \\
${\rm J1224-6407}$  &  $   -60.02$  &  $  -1.42$  &  $  1.90$  &  $ 0.002$  &  $ 0.10$  &  $ 0.50$  &  $ 0.05$  &  $ 0.10$  &  $30.00$  \\
${\rm J1453-6413}$  &  $   -44.27$  &  $  -4.43$  &  $  1.50$  &  $ 0.005$  &  $ 0.10$  &  $ 0.25$  &  $ 0.10$  &  $ 0.10$  &  $46.00$  \\
${\rm J1600-5044}$  &  $   -29.31$  &  $   1.63$  &  $  1.50$  &  $ 0.001$  &  $ 0.10$  &  $ 1.00$  &  $ 0.10$  &  $ 0.10$  &  $61.00$  \\
${\rm J1600-5044}$  &  $   -29.31$  &  $   1.63$  &  $  3.60$  &  $ 0.001$  &  $ 0.10$  &  $ 0.50$  &  $ 0.20$  &  $ 0.20$  &  $61.00$  \\
${\rm J1559-4438}$  &  $   -25.46$  &  $   6.37$  &  $  1.30$  &  $ 0.020$  &  $ 0.00$  &  $ 0.90$  &  $ 0.07$  &  $ 0.07$  &  $64.54$  \\
${\rm J1709-4428}$  &  $   -16.90$  &  $  -2.68$  &  $  1.50$  &  $ 0.010$  &  $ 0.10$  &  $ 0.40$  &  $ 0.10$  &  $ 0.10$  &  $73.00$  \\
${\rm 1757-24}$  &  $     5.31$  &  $   0.02$  &  $  3.00$  &  $ 0.035$  &  $10.00$  &  $ 1.20$  &  $ 0.03$  &  $ 0.10$  &  $95.31$  \\
${\rm 1759-2205}$  &  $     7.47$  &  $   0.81$  &  $  2.00$  &  $ 0.055$  &  $ 1.30$  &  $ 1.20$  &  $ 0.02$  &  $ 0.10$  &  $97.47$  \\
${\rm 1821-24}$  &  $     7.80$  &  $  -5.58$  &  $  1.50$  &  $ 0.005$  &  $ 0.10$  &  $ 0.40$  &  $ 0.20$  &  $ 0.20$  &  $ 0.00$  \\
${\rm 1534+12}$  &  $    19.85$  &  $  48.34$  &  $  0.35$  &  $ 0.004$  &  $ 0.00$  &  $ 0.20$  &  $ 0.20$  &  $ 0.30$  &  $-70.15$  \\
${\rm Interarm2-3}$  &  $    31.00$  &  $   0.00$  &  $  4.20$  &  $ 0.010$  &  $ 0.10$  &  $ 0.80$  &  $ 0.45$  &  $ 0.20$  &  $-25.00$  \\
${\rm 1859+03}$  &  $    37.21$  &  $  -0.64$  &  $  6.00$  &  $ 0.100$  &  $ 1.00$  &  $ 0.90$  &  $ 0.02$  &  $ 0.04$  &  $127.21$  \\
${\rm Interarm3-4}$  &  $    45.00$  &  $   0.00$  &  $  2.00$  &  $ 0.024$  &  $ 0.10$  &  $ 0.80$  &  $ 0.30$  &  $ 0.20$  &  $-25.00$  \\
${\rm 1915+13}$  &  $    48.26$  &  $   0.62$  &  $  3.00$  &  $ 0.024$  &  $20.00$  &  $ 0.20$  &  $ 0.20$  &  $ 0.20$  &  $-41.74$  \\
${\rm 2334+61}$  &  $   114.28$  &  $   0.23$  &  $  2.50$  &  $ 0.005$  &  $ 0.10$  &  $ 0.50$  &  $ 0.10$  &  $ 0.10$  &  $24.00$  \\
${\rm 0138+59}$  &  $   129.15$  &  $  -2.10$  &  $  1.50$  &  $ 0.017$  &  $ 0.10$  &  $ 0.70$  &  $ 0.10$  &  $ 0.10$  &  $39.00$  \\

\\
\enddata
\tablenotetext{~}{The columns are 
(1) line of sight name;
(2)-(3) Galactic coordinates in degrees;
(4) void distance (kpc);
(5) void electron density (cm$^{-3}$);
(6) void fluctuation parameter;
(7)-(9) ellipsoidal semi-axes (kpc);
(10) rotation angle (degrees) of semi-major axis about the $z$ axis,
referenced to the $x$ axis.
}
\end{deluxetable}


\clearpage
\appendix 
\section{A. LISM Model Form}\label{app:LISM}

The local ISM is modeled with four components that augment
the large scale Galactic features. These are
the local hot bubble (LHB) in which the Sun sits, a low-density
region (LDR) primarily in the first quadrant of the Galaxy,
a local superbubble (LSB) region in the third quadrant, and
the ``Loop~I'' feature that includes the North Polar Spur 
(e.g, Berkhuijsen, Haslam, \& Salter 1971; Spoelstra 1972) 

We model the LDR and LSB as ellipsoids with one axis perpendicular to
the plane of the Galaxy and the other making an angle $\theta$ with
the $x$ direction (i.e. $\ell = 90^{\circ}$).   The semi-major axes
are denoted $a,b,c$ and  the ellipsoid center
is given by $\xbar, \ybar, \zbar$. 
The internal density $n_e$ and fluctuation parameter $F$ are also
individual attributes for each region.  Parameter values are
given in Table \ref{tab:lism}. 

For the LHB we use the work of Bhat \etal (1999), Sfeir \etal (1999),
Vergely \etal (2001) and Ma\'iz-Apell\'aniz (2001) as a guide
for defining its structure.   Using contours of \ion{Na}{1} absorption
given by Sfeir \etal (1999), which delineate the absence of absorbing gas,
and the electron density implied by X-ray observations 
(Snowden \etal 1998), $n_e \sim 0.005$ cm$^{-3}$, we define the
LHB as follows.    The LHB is a slanted, ellipsoidal  cylinder
with cross section in the $x-y$ plane described by the parameters
$a$ and $b$.     The cylinder has total length $2c$ in the 
$z$ direction and a mean $z$ given by $\zbar$.   
For $z\ge 0$, the cylinder has constant cross section at constant $z$.
For $z<0$, $a$ declines linearly, reaching  zero at $z  = \zbar - c$.
the $b$ parameter is constant for all $z$ within the cylinder.
Finally, the cylinder axis is slanted in the y-z plane
with angle $\tan\theta = dy/dz$.    The internal density
and fluctuation parameter are also parameters for the LHB. 
Parameter values  are given in Table \ref{tab:lism}.

The Loop I component is modeled only for $z\ge 0$ 
as a hemisphere  of radius $r$ 
surrounded by a shell of thickness $\Delta r$.  
The hemispherical volume and shell
have different internal densities and F values. 
Parameter values given in Table \ref{tab:lism}
are guided by those given by Heiles (1998) but result from
fitting  for the best values of the parameters.

All four LISM regions have internal electron densities that are
less than those in the large-scale Galactic structures.    
Therefore, to ensure that densities are mutually exclusive (not additive)
between different components, we combine components
by assigning weight factors
$\wlhb, \wlsb$, $\wldr$ and $\wloopI$ to each region and, together, use
them to define an overall LISM weight, $\wlism$.
The weights are either 0 or 1.   
They are defined with the following hierarchy:
LHB:LoopI:LSB:LDR:ISM, meaning that 
the LHB overrides all other components (LISM and large scale),
Loop I overrides the LSB,  
the LSB overrides the LDR, and any LISM component overrides
the large scale components.     For the LDR and LSB, the weight is
unity inside the $e^{-1}$ contour of each ellipsoid while for the
LHB, the weight is unity anywhere inside the cylinder.
For Loop I, the weight is unity anywhere inside the hemispherical sphere
or shell.

\section{B. Fortran Code }\label{app:fortran}
 
Our model is implemented in Fortran routines similar in functionality
to those presented in TC93, but that represent a complete revision
according to the new features presented in the main text.  Copies of
the source and input files are available over the Internet at
\url{http://www.astro.cornell.edu/\~cordes/NE2001} and
\url{http://rsd-www.nrl.navy.mil/7213/lazio/ne\_model/}.  The code is
packaged as a \texttt{tar} file, {\tt NE2001.tar}, that includes a
makefile for compiling the code in a Unix/Linux environment.

The makefile can compile the code into either an interactive program
(\texttt{make pgm}) or (static) library (\texttt{make lib}).  The
interactive program {\tt NE2001} evaluates the model by returning the
integrated measures (DM, SM, etc.)  and/or distance given an input
direction and DM or distance.  The library \texttt{libNE2001.a}
allows one to incorporate the Fortran routines into user-written
programs.  Compiling the library into a program would be in a manner similar
to the following:
\begin{center}
\mbox{f77 -L/path/to/library/directory program.f -lNE2001.}
\end{center}

Integrations are performed in the the subroutine {\tt dmdsm}, which
evaluates the model by making calls to subroutine {\tt
density\_2001}.  The call to {\tt dmdsm} is of the form 

\begin{center}
\mbox{\tt call dmdsm(l,b,ndir,dm,dist,limit,sm,smtau,smtheta,smiso)} .
\end{center}
Here the input data include Galactic longitude and latitude,
 {\tt l} and {\tt b} (in radians), and a flag {\tt ndir} indicating whether
distance is to be calculated from dispersion measure ({\tt ndir}$\ge
0$), or {\em vice-versa} ({\tt ndir}$<0$).  In either case, {\tt dm}
and {\tt dist} have units of pc cm$^{-3}\,$ and kpc, respectively.  A
flag {\tt limit} is set if {\tt ndir}$\ge 0$ and the model distance is
a lower limit; this will occur, for example, if a large {\tt dm} is
specified at high Galactic latitude.  The subroutine also returns
four estimates of scattering measure, all having units
kpc m$^{-20/3}$.  The first, {\tt sm}, conforms
to the definition in Eq.~(\ref{eq:smdef}) with uniform weighting along
the line of sight.  The next two estimates, {\tt smtau} and {\tt
smtheta}, correspond to line-of-sight weightings appropriate for
temporal and angular broadening of Galactic sources, respectively.
Temporal broadening emphasizes scattering material midway between
source and observer, while angular broadening favors material closest
to the observer; see Eqs.~(A14, B2) of Cordes, Weisberg, \& Boriakoff
(1985). The last estimate, {\tt smiso}, uses the weighting appropriate
for calculating the isoplanatic angle of scattering.  

Integrations in {\tt dmdsm} involve evaluations of the model
at a given Galactic location $(x,y,z)$ through a call to
subroutine {\tt density\_2001},
where {\tt x,y,z} are Galactocentric Cartesian coordinates, measured
in kiloparsecs, with the axes parallel to $(l,b)=(90^\circ,0^\circ)$,
$(180^\circ,0^\circ)$, and $(0^\circ,90^\circ)$:  

{\setlength{\parindent}{1.0in}
\begin{minipage}{\linewidth}
{\tt
\begin{verbatim}
      call density_2001(x,y,z,
     .  ne1,ne2,nea,negc,nelism,necN,nevN,
     .  F1, F2, Fa, Fgc, Flism, FcN, FvN,
     .  whicharm, wlism, wldr, wlhb, wlsb, wloopI,
     .  hitclump, hitvoid, wvoid).
\end{verbatim}
}
\end{minipage}
}
\noindent
The routine returns values for the electron density in seven components 
({\tt ne1, $\cdots$, nevN}), the corresponding $F$ parameters
\hbox{({\tt F1, $\cdots$, FvN})}, followed by a series of 
integer-valued flags.   The meanings of these flags
are as follows: 
\begin{enumerate}
\item {\tt whicharm = 0, $\cdots$, 5} indicates which spiral arm contributes to the density, with numbering as in the text and where a zero value denotes an
interarm region.
\item
{\tt wlism, wldr, wlhb, wlsb} and {\tt wloopI} take on values
of 0 or 1 as described in Appendix~\ref{app:LISM}.
\item
{\tt hitclump} denotes whether a clump has been intersected in the integration; 
if so,
then {\tt hitclump} denotes the clump number in the table of
clumps; if not, {\tt hitclump = 0}. 
\item 
{\tt hitvoid} works in the same fashion for voids; additionally,
{\tt wvoid = 0,1} is used in evaluating the total density and indicates
if a void has been hit ({\tt wvoid = 1}). 
\end{enumerate}

The calling program,  {\tt NE2001}, is executed using command-line
arguments that specify the Galactic longitude and latitude,
an input DM or distance value, and a flag ({\tt ndir}) that
specifies whether a distance is calculated from DM
({\tt ndir $\ge$ 0}) or a DM calculated from an input 
distance ({\tt ndir $<$ 0}):

{\setlength{\parindent}{1.0in}
\begin{minipage}{\linewidth}
{\tt
\begin{verbatim}
 Usage: NE2001 l b DM/D ndir
        l (deg)
        b (deg)
        DM/D (pc cm^{-3} or kpc)
        ndir = 1 (DM->D)   -1 (D->DM) 
\end{verbatim}
}
\end{minipage}
}

Program {\tt NE2001} uses output from {\tt dmdsm} to calculate
scattering and scintillation quantities by making suitable calls
to a series of functions.   Input distances, scattering
measures, frequencies and velocities are in standard units
(kpc, kpc m$^{-20/3}$, GHz, and km s$^{-1}$):
\begin{enumerate}
\itemsep -2pt
\item 
{\tt tauiss(d,sm,nu)}: calculates the pulse broadening time,
$\taud$ (ms).
See Eq.~\ref{eq:taud}.
\item 
{\tt scintbw(d,sm,nu)}: calculates the scintillation bandwidth,
$\dnud$ (MHz).
See Eq.~\ref{eq:dnud}.
\item 
{\tt scintime(sm,nu,vperp)}: calculates the scintillation time,
$\Delta t_{\rm ISS}$ (sec) (Cordes \& Lazio 1991; Cordes \& Rickett 1998).
\item 
{\tt specbroad(sm,nu,vperp)}: calculates the spectral broadening,
$\Delta\nu_{\rm b}$
 (Hz), that is proportional to the reciprocal of the scintillation time
 (Cordes \& Lazio 1991).
\item 
{\tt theta\_xgal(sm,nu)}: calculates the angular broadening,
$\theta_d$ (mas), appropriate for the scattering geometry for
an extragalactic source.  See Eq.~\ref{eq:thetad}.
\item 
{\tt theta\_gal(sm,nu)}: calculates the angular broadening,
$\theta_d$ (mas), of a Galactic source.
See Eq.~\ref{eq:thetad}.
\item 
{\tt em(sm)}: calculates the emission measure, \EM\ (pc cm$^{-6}$),
 associated with the scattering measure; note that the value calculated
assumes a particular outer scale for a Kolmogorov wavenumber spectrum
and represents a lower bound on \EM\ (see Eq.~\ref{eq:sm2em}).
\item 
{\tt theta\_iso(smiso,nu)}: calculates the isoplanatic
angle, $\theta_{\rm iso}$ (mas), 
the region on the sky over which scintillations are correlated. 
\item 
{\tt transition\_frequency(sm,smtau,smtheta,dintegrate)}: 
calculates the frequency of transition, $\nutrans$ (GHz),  
between the weak and strong scattering regimes (Eq.~\ref{eq:nutrans}).
\end{enumerate}

Sample output for $\ell = 45^{\circ}, b = 5^{\circ}$, 
and \DM\ = 50 pc cm$^{-3}$ is:

{\setlength{\parindent}{1.0in}
\begin{minipage}{\linewidth}
{\tt
\begin{verbatim}
#NE2001 input: 4 parameters
   45.0000         l         (deg)                    GalacticLongitude
    5.0000         b         (deg)                    GalacticLatitude
   50.0000         DM/D      (pc-cm^{-3}_or_kpc)      Input_DM_or_Distance
         1         ndir      1:DM->D;-1:D->DM         Which?(DM_or_D)
 #NE2001 output: 14 values
    2.6365         DIST      (kpc)                    ModelDistance
   50.0000         DM        (pc-cm^{-3})             DispersionMeasure
    4.3578         DMz       (pc-cm^{-3})             DM_Zcomponent
0.3528E-03         SM        (kpc-m^{-20/3})          ScatteringMeasure
0.2367E-03         SMtau     (kpc-m^{-20/3})          SM_PulseBroadening
0.7719E-04         SMtheta   (kpc-m^{-20/3})          SM_GalAngularBroadening
0.1307E-02         SMiso     (kpc-m^{-20/3})          SM_IsoplanaticAngle
0.1921E+00         EM        (pc-cm^{-6})             EmissionMeasure_from_SM
0.1293E-03         TAU       (ms)                     PulseBroadening @1GHz
0.1428E+01         SBW       (MHz)                    ScintBW @1GHz
0.4943E+03         SCINTIME  (s)                      ScintTime @1GHz @100 km/s
0.2420E+00         THETA_G   (mas)                    AngBroadeningGal @1GHz
0.1086E+01         THETA_X   (mas)                    AngBroadeningXgal @1GHz
     14.02         NU_T      (GHz)                    TransitionFrequency
\end{verbatim}
}
\end{minipage}
}

The Fortran program can be run using a perl script (also included
in the tar file) which allows selection of
an individual field for output:

{\setlength{\parindent}{1.0in}
\begin{minipage}{\linewidth}
{\tt
\begin{verbatim}
run_NE2001.pl 
Usage:
        run_NE2001      l    b       DM/D         ndir    field 
                       deg  deg    pc-cm^{-3}     1,-1    D etc 
                                   or kpc 
        Possible Fields (case insensitive):
        Dist, DM, SM, EM, TAU, SBW, SCINTIME, THETA_G, THETA_X, NU_T, ALL
\end{verbatim}
}
\end{minipage}
}

\clearpage

\end{document}